\documentstyle[12pt,a4]{article}

\def\<{\langle}
\def\>{\rangle}
\def\/{\sqrt{}}
\def\a{\alpha}
\def\b{\beta}
\def\d{\delta}
\def\l{\lambda}
\def\g{\gamma}
\def\e{\epsilon}
\def\k{\kappa}
\def\s{\sigma}
\def\th{\tilde{h}}
\def\tf{\tilde{f}}
\def\tg{\tilde{\g}}
\def\tl{\tilde{\l}}
\def\ts{\tilde{\s}}
\def\tw{\tilde{w}}
\def\gc{\g^{{}_{(C)}}}
\def\gs{\g^{{}_{(S)}}}
\def\gv{\g^{{}_{(V)}}}
\def\gt{\g^{{}_{(T)}}}
\def\bs{\b^{{}_{(S)}}}
\def\bv{\b^{{}_{(V)}}}
\def\bt{\b^{{}_{(T)}}}
\def\gcdot{\dot{\g}^{{}_{(C)}}}
\def\gsdot{\dot{\g}^{{}_{(S)}}}
\def\gvdot{\dot{\g}^{{}_{(V)}}}
\def\gtdot{\dot{\g}^{{}_{(T)}}}

\begin{document}

\vskip 1.5 cm
\centerline{\Large
Evolution of the discrepancy between a universe and its model}
\vskip 1.0 cm

\centerline{\large {\sc Masafumi Seriu}}
\vskip .2cm
\centerline{\it Institute of Cosmology}
\centerline{\it Department of Physics \& Astronomy}
\centerline{\it Tufts University}
\centerline{\it Massachusetts 02155, USA}
\centerline{*}
\centerline{\it  Department of Physics, Fukui University}
\centerline{\it  Fukui 910-8507, Japan~\footnote{
Present address. E-mail: mseriu@edu00.f-edu.fukui-u.ac.jp}}
\vskip 1.5cm

\begin{abstract}
We study a fundamental issue in cosmology: Whether we can rely on
a cosmological model  to understand the real history of the Universe.
This fundamental, still unresolved issue is often called
the ``model-fitting problem (or averaging problem) in cosmology''.
Here we analyze  this issue with the help of
the spectral scheme prepared in the preceding studies.

Choosing two specific spatial geometries ${\cal G}$ and ${\cal G}'$ that are
very close to each other,
we investigate explicitly the time evolution of the spectral
distance between them;  as two spatial geometries  ${\cal G}$ and ${\cal G}'$,
we choose a flat 3-torus  and a perturbed geometry around it, mimicking
the relation of  a ``model universe'' and the ``real Universe''.
 Then we estimate  the spectral distance between them,
 $d_N ({\cal G},{\cal G}')$,
and investigate  its time evolution explicitly.
This analysis is done efficiently by making use of
the basic results of the standard linear structure-formation theory.

We observe that, as far as the linear approximation of the geometrical
perturbation is valid, $d_N ({\cal G},{\cal G}')$ does not increase
with time prominently,
rather it shows the tendency to decrease.
This result is compatible with the general belief in  the reliability  of
 describing the Universe by means of  a model, and
 calls for more detailed studies along the same
line including the investigation of  wider class of spacetimes and
the analysis beyond the linear regime.
\end{abstract}

\section{Introduction}
\label{section:I}

We study in this paper one of the fundamental issues in cosmology:
How precisely   a  cosmological model traces
 the time evolution of the real Universe.
This significant problem has been raised and studied for a long time
but we have not yet caught a glimpse of
 a satisfactory understanding of this issue.
 It is often called the ``model-fitting problem in
 cosmology" or the ``averaging problem in cosmology"~\cite{AVE,AVE2}.

This problem originates from  the very nature of cosmology.
 Cosmology  is our attempt to  grasp our Universe
 by reference to  a model. The real  Universe is full of complexity while
 a model, which serves as a format of perceiving reality,
 is much simpler than the reality itself.
In cosmology, thus,  there is in principle an inevitable
discrepancy  between the reality and its model.

When we try to guess the past and future
 of the Universe based on its model, thus,  we need a guarantee
 that the inevitable discrepancy
 between the reality and its model does not  develop so much within the time
 scale of concern.  Considering
 the highly nonlinear nature of the Einstein equation, the enlargement of
 the discrepancy would not be a very surprising result even if it
 were the case.   Therefore
this issue  should be thoroughly clarified before we state anything
meaningful for the past or future of our Universe.
One might symbolically state the problem depicted here  as
``{\it Is cosmology  possible?}"

The central difficulty of dealing with the issue of inevitable discrepancy
may arise from the following two facts:
\begin{description}
\item{\it (D1)} It is difficult to define a reasonable
  procedure of ``averaging geometry" in a self-consistent manner.
  For instance, it is difficult to define the averaged spatial metric
  $\< h_{ab} \>$ in a spatial-diffeo-invariant manner.
\item{\it (D2)}
 On one hand, we need to handle
two or more  universes  (viz. the real one and  its models) at the same time;
on the other hand, there has been no efficient mathematical language for
comparing two or more geometries with each other.
\end{description}

At first sight, the difficulties {\it (D1)} and {\it (D2)} look  mutually independent.
When we contemplate on them deeper, however, we realize that
they are both rooted in the issue of how to quantify an intuitive statement
``the two shapes look similar": The issue {\it (D1)} is reduced to the issue of
how to relate the averaged geometry to the original geometry in a reasonable
manner. The issue {\it (D2)} is about the comparison procedure for two given
geometries.   Extracting the essence from {\it (D1)} and {\it (D2)}, then,
we infer that  the concept of ``closeness" between two
geometries would play a key role.  We pursue this viewpoint further in
this paper, making use of our preceding related studies.

In preceding work
we have indeed prepared a framework
 in which the measure of closeness between geometries plays a central
 role~\cite{MS-spectral,MS-space,MS-evolution}, and
 have made some preliminary investigations on the
model-fitting problem based on this framework~\cite{MS-AVE,MS-JGRG}.
(For simplicity we call the framework {\it spectral scheme} hereafter).
The purpose of this paper is to make
a more explicit investigation by the spectral scheme
on the time evolution of the discrepancy between
two nearby geometries, $\cal G$ and ${\cal G}'$,
by choosing concrete, tractable geometries for $\cal G$ and ${\cal G}'$.

Before going into details, let us first discuss
why handling the  discrepancy between two universes is
so difficult  and how we try to tackle this problem by introducing
the spectral scheme.

For this purpose the following example may clearly illustrate the situation.
Let ${\cal G}_0$ be  some topologically complicated space,
 with various small topological handles attached~\cite{MS-scale,MS-spectral}.
When we observe ${\cal G}_0$ at the energy scale $E$, the handles smaller
than  $E^{-1}$ would not be observed, resulting in a  simpler
effective geometry, say ${\cal G}_E$.
If the energy scale is decreased further to $E'$ ($E > E'$),
the handles smaller than  ${E'}^{-1}$ effectively disappear,
resulting in a much  simpler effective geometry ${\cal G}_{E'}$.
A similar phenomenon occurs when we change observational apparatus,
instead of changing $E$;
a portion of geometrical information
incompatible with the observational apparatus
would in effect disappear so that  the resulting effective geometry depends on
the apparatus used.

In this way, we are led to  the concept of
{\it effective geometry}, viz. geometrical structure
varying as a ``function" of observational energy $E$
 and observational apparatus $A$, which may be indicated
 by ${\cal G}_{(E,A)}$.
We intuitively know that
 ${\cal G}_{(E,A)}$ is effectively ``close" to ${\cal G}_0$ and that
 ${\cal G}_{(E,A)}$ can serve as a model for ${\cal G}_0$ when we
 are interested in  the phenomena labeled by $(E,A)$.

On the other hand, ${\cal G}_{(E,A)}$ (for some $E$ and $A$) is  totally
different from the original geometry ${\cal G}_0$ from the viewpoint of
traditional mathematical theories of geometry and topology.
Even though  ${\cal G}_{(E,A)}$ and  ${\cal G}_0$ are physically ``close"
to each other,
mathematics classifies them into different topology classes.
Topology is a scale-free concept and the scale of a handle is not counted,
while in spacetime physics it  should also  be
taken into account because the observational energy scale enters into
the argument~\cite{MS-scale,MS-spectral}.

The central difficulty of analyzing the issue of discrepancy is, then, that
we are lacking in  a scheme of geometrical approximation which
makes it possible to regard ${\cal G}_{(E,A)}$ as an approximation (a ``model")
 of ${\cal G}_0$, even though they are totally different in a
 conservative sense.
 Thus the first thing to do is to construct a new framework which justifies
 our intuition that ${\cal G}_{(E,A)}$ is  ``close" to ${\cal G}_0$
 in a certain sense. In this way we return to the issue of
 ``closeness" discussed after {\it (D1)} and {\it (D2)}.

With the above consideration, it is now clear that the following points
should be clarified;
\begin{description}
\item{(a)} How to quantify ``discrepancy" or ``closeness" between
         two geometries.
\item{(b)} How to take into account the
   ``coarseness" of  truncating information of geometry.
   (Scale-dependence of effective geometry.)
\item{(c)} How to take into account the type of apparatus used
for collecting information of geometry.
   (Apparatus-dependence of effective geometry.)
\end{description}

 It has been  shown that a measure of closeness
between two geometries, say ${\cal G}$ and ${\cal G}'$,
can be defined in terms of the spectra~\cite{MS-spectral};
here the term  ``spectra"  means
a sequence of eigenvalues of a certain elliptic Hermitian
operator, numbered in an increasing order.
(We mainly consider  the Laplacian  $\Delta$ as a typical
Hermitian  elliptic operator though the basic idea itself is  universal.)
The measure of closeness is defined  by comparing the spectra
for ${\cal G}$ with those for  ${\cal G}'$ ({\it Issue (a)}).
Furthermore, only the first $N$ spectra ($\l_1, \l_2, \cdots, \l_N$) are
compared, and in this manner, a natural cut-off scale $\l_N^{-1/2}$ is
purposefully introduced.
 This procedure  can be symbolically described
 as ``comparing the sound of ${\cal G}$ with the
 sound of ${\cal G}'$ ".

First of all, the spectra are the spatial-diffeomorphism invariant so that
they are desirable quantities for  quantifying  ``closeness" (see {\it (D1)}).
Next, on dimensional grounds,
the lower (higher) spectra  reflect the larger (smaller)
scale properties of the geometry. By comparing only the low-lying spectra
up to the $N$-th spectrum $\l_N$, thus,
${\cal G}$ and ${\cal G}'$ are compared in a coarse-grained manner,
neglecting the difference smaller than the cut-off scale
corresponding to $\l_N$.
Thus the measure of closeness $d_N ({\cal G}, {\cal G}') $
 introduced in this way  naturally takes care of the  scale-dependent nature
 of spatial geometry ({\it Issue (b)}).
Finally, the choice of the Hermitian
operator determines the type  of vibration modes  used
for probing  geometry, so that in principle it corresponds to the
observational apparatus used for measuring geometry. Thus the choice of the
operator defines which  aspects of geometry are compared to
 measure  the closeness between the spaces. As a result,
  our  measure of closeness becomes apparatus-dependent, too ({\it Issue (c)}).

In this manner, we are now capable of  quantifying  the  ``closeness"
between  two given geometries as
a ``function" of the  coarse-graining scale and
the observational apparatus~\cite{MS-spectral}.
For a given geometry ${\cal G}'$,  the
{\it optimal model} of ${\cal G}'$ among a set of models can be defined as
 the model geometry $\cal G$ closest to ${\cal G}'$ when  measured by $d_N$.
Therefore we realize that
a suitable model for a given ``reality" ${\cal G}'$ is not unique,
but it varies in accordance with  the observational scale and
apparatus~\cite{MS-AVE,MS-JGRG}.
The mapping ${\cal G}' \mapsto {\cal G}$ here may be interpreted as
``averaging the geometry ${\cal G}'$" if one wishes. The unique feature of this
approach is, however, it does not resort to the ambiguous averaging procedures
such as defining the ``averaged metric" $\< h_{ab} \>$; rather the whole
of the geometry is directly mapped to some suitable model geometry.

Furthermore, it has been shown that the space of all spaces ${\cal S}_N$,
which is a set of all compact  Riemannian geometries equipped with
the measure of closeness $d_N$, forms a metrizable space~\cite{MS-space}.
This result makes it justified to call  the measure of closeness
$d_N ({\cal G}, {\cal G}')$
as the {\it spectral distance} between ${\cal G}$ and  ${\cal G}'$.
At the same time, it means that
 we have constructed a basic arena ${\cal S}_N$ for analyzing
 fundamental problems in spacetime physics,  such as the one discussed
 in the present paper.

One of the advantages of the spectral scheme is that
the basic quantities $\{ \l_n  \}$ carry clear, well-defined meaning
 both physically and mathematically.
 Physically they represent the vibration modes of the space
when ``tapped'' by a certain apparatus ($\Delta$); mathematically they are
the eigenvalues of a certain elliptic operator defined on a space  and
their investigation forms one active subject in geometry.

In reality, it is true that only a restricted number of  cases are
known for which whole of the spectra, $\{ \l_n  \}_{n=1}^\infty$,
can be obtained explicitly.
Even so, it is still meaningful to construct a general framework
based on $\{ \l_n  \}$:
First, let us recall that we do not need the whole of the spectra
but  only  the first $N$ spectra
($\l_1, \l_2, \cdots, \l_N$) to be compared.
It is  known that the lower spectra are
relatively easier to obtain by, for instance, the numerical
calculations~\footnote{Indeed there are already several numerical
calculations~\cite{Numerical}
(based on the finite element method) which
provide the first 15 spectra or so for the case of 3-dimensional hyperbolic
geometries with non-trivial topologies.
This kind of numerical techniques are certain to
develop in the near future.}
as well as analytical methods~\cite{CH}, and
these lower spectra  contain enough information on the {\it global}
nature of the space.  Second, in many cases,
we are more interested in  the difference of the spectra ($\d \l_n$)
between two geometries (or between two neighboring time-slices in one
spacetime) than the spectra themselves, and usually
we can estimate  the former even when the latter are not explicitly available.
Third, it is  meaningful that the spectra are theoretically well-defined and
{\it in principle} determined, even though not determined explicitly in many
cases.~\footnote{
The situation here is very similar to the one for quantum field theory
on a curved spacetime.
There are only restricted number of spacetimes available for which
we can explicitly obtain the complete set of positive frequency solutions
of the field equation. However the general framework of quantum field theory
is still worth constructing.}
Finally, the situation is not proper to the spectral scheme;
whatever we choose as a framework of analysis,
investigations  of complicated geometries  do not go straightforwardly anyway.

It is now possible to  picture the situation in question as follows:
Each point in the space  ${\cal S}_N$ indicates a certain
geometry  viewed with the apparatus $\Delta$
at the scale $\l_N$.\footnote{More rigorously,
it indicates a  class of geometries that look similar to each other
when viewed with the apparatus $\Delta$
at the scale $\l_N$.}
 A set of models is represented by the points distributed throughout
 ${\cal S}_N$. The optimal model $\cal G$ is then represented  by the point
 among them closest to the point representing the reality
 ${\cal G}'$~\cite{MS-AVE}.
Thus, the real geometry ${\cal G}'$ and its
model geometry $\cal G$ are represented by two points
 that are very close to each other in  ${\cal S}_N$.
 These two points move in time and follow two trajectories in ${\cal S}_N$,
 and the relative behavior of these two trajectories is crucial for
 a model $\cal G$ to be a good approximation of ${\cal G}'$.
What is essential is, then, to investigate
 the time evolution of the spectral distance $d_N({\cal G},{\cal G}')$.
 For this purpose,
 the evolution  equations for the spectra of the universe is required  since
the spectral distance is defined solely in terms of the spectra.
The spectral evolution equations have been indeed derived\cite{MS-evolution},
and they  can be regarded as the spectral representation
of the Einstein equation.
 Now we have constructed  a formalism which consists of the following triad;
(I) the spectral distance  $d_N$,
  (II) the space of all spaces ${\cal S}_N$ and (III) the spectral evolution
  equations.

In this paper, we apply
 the spectral scheme explicitly to a concrete, tractable situation
 to get deeper understanding of
  the relation between the reality and its model.
We construct  a tractable pair of spaces imitating the relation between
 the real Universe and its model,
 and estimate the time evolution of their spectral distance
explicitly. As the first investigation in this direction,
we here investigate only the linear-regime, viz.
 the period  when the discrepancy can be
 regarded as small. Though the situation analyzed here is inevitably
  limited, the present  analysis would give us insight for
 the cases when the discrepancy is larger than the present case
 and would  provide a motivation for further studies for these cases.

In Section \ref{section:II}, we will summarize  the basic formulas
in the spectral scheme only to a necessary extent
for the analysis in this paper.
In Section \ref{section:III}, we construct
two spacetimes, mimicking the relation between
the Universe and its model; as for their spatial sections,
a flat 3-torus is chosen for  what is regard as
 a ``model universe'' (${\cal G}$), while
a perturbed geometry around ${\cal G}$
is chosen for what plays the role of  the ``real Universe''
 (${\cal G}'$). Then we prepare
several fundamental quantities and formulas for these spacetimes
 in terms of the spectral scheme.
We combine the standard linear-perturbation theory with
the spectral scheme in Section \ref{section:IV},
and derive the time-dependence of various
coefficients necessary for our analysis.
With these preparations, we estimate
the time-development  of the spectral distance between
${\cal G}$ and ${\cal G}'$  in Section \ref{section:V} (the main results
are Eqs.(\ref{eq:final-result1})-(\ref{eq:final-result4})).
Section \ref{section:VI} is devoted for several discussions.

\section{Fundamental formulas in the spectral scheme}
\label{section:II}

Let us recall some results in the spectral scheme  necessary for
later analysis.\footnote{
The reader is advised to refer to Refs. \cite{MS-spectral},
\cite{MS-space} and \cite{MS-evolution} for more details.}

Let us consider a geometry ${\cal G}$ described by
a $(D-1)$-dimensional compact
Riemannian manifold without boundaries, $(\Sigma, h)$.\footnote{
In this section, $D$ indicates the dimension of a spacetime, so that
its spatial section becomes $(D-1)$-dimensional.}

 We consider an eigenvalue problem of the Laplacian
 $\Delta$ on $(\Sigma, h)$,
$\Delta f = -\l f$. Then we get the {\it spectra},
viz. the sequence of eigenvalues  arranged in an increasing order,
 $\{ \l_n \}_{n=0,1,2,\cdots}$
$:= \{ 0=\l_0 < \l_1 \leq \l_2 \leq \cdots \leq \l_n$$
\leq \cdots \}$.
At the same time, we get the set of real-valued eigenfunctions corresponding to
the spectra,   $\{ f_n  \}_{n=0,1,2,\cdots}$,  normalized as
$(f_m,\ f_n):=\int_{{}_\Sigma}\ f_m \ f_n \  \/ = \d_{mn}$ ,
where the natural integral measure on $(\Sigma, h)$
is indicated  by $\sqrt{}:=\sqrt{\det (h_{ab})}$.

Now we consider two geometries
${\cal G}$ and  ${\cal G}'$.
Let $\{\l_m \}_{m=0}^\infty$ and  $\{\l'_n \}_{n=0}^\infty$ be
 the spectra for  ${\cal G}$ and  for ${\cal G}'$, respectively.
Then  the spectral measure of closeness between $\cal G$ and ${\cal G}'$
 of order $N$, $d_N ({\cal G},{\cal G}')$,  is defined
 as~\cite{MS-space,MS-AVE}
\begin{equation}
  d_N ({\cal G},{\cal G}'):= \sum_{n=1}^N {\cal F}
  \left( \frac{\l'_n}{\l_n} \right)\ \ ,
\label{eq:d_N_general}
\end{equation}
  where ${\cal F}(x)$ ($x>0$) is any continuous function  satisfying
\begin{eqnarray}
  {\cal F}(1)&=&0 \ \ , \nonumber \\
  {\cal F}(y)&>&{\cal F}(x)\ \  {\rm for} \ \   y > x \geq 1 \ \ , \nonumber \\
  {\cal F}(1/x)&=&{\cal F}(x)\ \ .
\label{eq:F_cond}
\end{eqnarray}
  For practical applications, it is convenient
  to  assume further $(i)$ ${\cal F}(x)$ is smooth at $x=1$
  (then  ${\cal F}' (1)=0$ and
  ${\cal F}''(1) \geq 0$ from the conditions (\ref{eq:F_cond}))
  and $(ii)$  ${\cal F}''(1)>0$. The  postulation $(ii)$ is
    for making $d_N ({\cal G},{\cal G}')$ sensitive enough to detect a fine
    difference between ${\cal G}$ and ${\cal G}'$ when
  they are very close to each other
  (see Eq. (\ref{eq:d_N-close})).

It is convenient to set   ${\cal F}$ to be
 ${\cal F}_1(x)=\frac{1}{2} \ln \frac{1}{2}(\sqrt{x}+1/\sqrt{x})$.
Then  Eq.(\ref{eq:d_N_general}) becomes~\cite{MS-spectral}
\begin{equation}
d_N ({\cal G},{\cal G}')
=\frac{1}{2} \sum_{n=1}^N \ln \frac{1}{2}
\left(
\sqrt{\frac{\l_n'}{\l_n}}
+\sqrt{\frac{\l_n}{\l'_n}}
\right)\ \ .
\label{eq:d_N}
\end{equation}

Now, let  $Riem$ be  the space
 of all $(D-1)$-dimensional, compact Riemannian geometries without
 boundaries. We then consider a space  $(Riem, d_N)/_\sim $, viz. the space
 $Riem$ equipped with the measure of closeness $d_N$ (Eq.(\ref{eq:d_N}))
 along with a natural identification ($\sim$) of
 isospectral manifolds.\footnote{When two Riemannian manifolds possess
 the identical spectra of the Laplacian even though they are not
 isometric, they are said to be {\it isospectral} to each other.
 There are some known examples of the isospectral manifolds.
 For more details, see Refs. \cite{CH} and \cite{Kac}.
 For the physical interpretation of isospectral manifolds, see
 Refs. \cite{MS-AVE}  and \cite{MS-JGRG}.
 }

 It is proved that the space of all geometries $(Riem, d_N)/_\sim $
 forms  a metrizable space~\cite{MS-space}.\footnote{
 It turns out that~\cite{MS-space} the distance function for metrization
 is provided by Eq.(\ref{eq:d_N_general}) with the choice
 ${\cal F}_0(x):=\frac{1}{2}\ln\max(\sqrt{x},1/\sqrt{x})$ for
 $\cal F$.
 } It means that
 the measure of closeness $d_N$ (Eq.(\ref{eq:d_N})) can be regarded as
 if it were a distance in spite of  its mild violation of
 the triangle inequality~\cite{MS-JGRG,MS-space}.
 From now on, thus, let us call
 $d_N({\cal G},{\cal G}')$ in Eq.(\ref{eq:d_N}) the {\it spectral distance}
 of order $N$ between two geometries $\cal G$ and ${\cal G}'$.
  Let ${\cal S}_N$ be the completion of $(Riem, d_N)/_\sim $
  by means of the latter's  metrizable structure, since
  such a complete space is mathematically more desirable.
    One might call
 ${\cal S}_N$ the {\it space of all spaces} of order $N$, or
 the {\it spectral space} of geometries of order $N$.
 Mainly due to  its metrizable nature, ${\cal S}_N$ possesses several
 nice properties such as  the second countability, paracompactness
 (then the partition of unity can be introduced on ${\cal S}_N$)  and
 locally-compactness (then an integral over ${\cal S}_N$ can
 be defined).\footnote{
 For basic facts of point set topology, see e.g. Refs.\cite{KE} and \cite{YA}.}

  When we focus on the discrepancy of two geometries that are
  very close to each other in ${\cal S}_N$, there is no significant
  difference  even if we replace ${\cal F}_1(x)$
  (used in Eq.(\ref{eq:d_N}))
   with some  other continuous function
   ${\cal F}(x)$ satisfying  the conditions in
    (\ref{eq:F_cond})(see Eq.(\ref{eq:d_N-close}) below and the arguments there
    on this point).

Let us introduce some convenient notations for later use.
Let   ${\cal A}$ and ${\cal A}_{ab}$ be  any function and any
symmetric tensor field, respectively,  on a spatial geometry
$(\Sigma , h)$. Then,
we define spatial-diffeomorphism invariant quantities
$\< {\cal A} {\>_{}}_n$ and $\< {\cal A}_{ab} {\>_{}}_n$   as
\[
\< {\cal A} {\>_{}}_n
  := \int_{\Sigma}\ f_n \  {\cal A} \ f_n \ \/\ , \ \
\< {\cal A}_{ab} {\>_{}}_n
       := \frac{1}{\l_n}\int_{\Sigma} \
            \ \partial^af_n \  {\cal A}_{ab} \ \partial^b f_n \ \/ \ \ ,
\]
where $f_n$ is the $n$-th eigenfunction.
For a quantity  $\< {\cal A}_{ab}{\>_{}}_n$,
it is assumed   $n \geq 1$.

Any function ${\cal A}(\cdot)$ on the spatial section
$\Sigma$ can be expanded  in terms of $\{ f_n \}_{n=0}^\infty $, such as
${\cal A}(\cdot)=\sum_{n=0}^\infty {\cal A}_n\ f_n(\cdot) \ \ $.
Here we note
\begin{equation}
 {\cal A}_n = (f_n\ , \ {\cal A}):=\int_{\Sigma} f_n \ {\cal A} \/ \ \ .
\label{eq:fourier-component}
\end{equation}
The $0$-th component ${\cal A}_0$ is related to the
spatial average of $\cal A$ over $\Sigma$,
${\cal A}_{\rm av}:=\frac{1}{V}\int_{{}_\Sigma} {\cal A} \/ $, as
\[
 {\cal A}_{\rm av}={\cal A}_0/\sqrt{V}\ \  ,
\]
where $V$ is the $(D-1)$-volume of the spatial section $(\Sigma , h )$.

 We also introduce the quantity
\begin{equation}
(l\ m\ n):= \<f_m {\>_{}}_{ln}=\int\ f_l\ f_m\ f_n \/ \ \ ,
\label{eq:(lmn)}
\end{equation}
which is totally symmetric in $l$, $m$ and $n$.

It is  useful to define $\e_{ab}$ and $r_{ab}$,
 the trace-free components  of the
extrinsic curvature $K_{ab}$  and the Ricci tensor
$\mbox{\boldmath $R$}_{ab}$ for $(\Sigma, h)$, respectively;
\begin{equation}
\e_{ab} := K_{ab}-\frac{1}{D-1}K h_{ab} \ \ ,\ \
r_{ab} :=
  \mbox{\boldmath $R$}_{ab}- \frac{1}{D-1}\mbox{\boldmath $R$}h_{ab} \ \ .
\label{eq:e/r}
\end{equation}
These quantities  describe anisotropy of the geometry $(\Sigma, h)$.

At this stage it is appropriate to make it clear
what the terms ``the real Universe"
(``reality") and ``the model" actually indicate in this paper. We shall
compare two spaces, $\cal G$ and ${\cal G}'$,
that are very close to each other in ${\cal S}_N$; this is
a totally well-defined situation without any ambiguity. Throughout this paper,
we deal with only  this mathematically well-posed problem.
Only on the final stage, however, we may interpret the results of
this analysis in the context of the averaging problem in cosmology,
supposing that one of the spaces corresponds to  the real Universe
and  the other to its model. To keep in mind this final interpretation,
 we symbolically refer to  the simpler space among the
 $\cal G$ and ${\cal G}'$ as the ``model",
while  the other space as the ``real Universe" (``reality").

It is now in principle possible to investigate the evolution of the
discrepancy between the reality and its model generally.
 In this paper, however, we restrict ourselves to the cases when
\begin{description}
\item{(A1)}  the discrepancy between $\cal G$ and ${\cal G}'$
              can be totally described in terms of  the
             difference  in the spatial metric, and
\item{(A2)}  they are initially so close to each other in ${\cal S}_N$
               that the linear treatment of the discrepancy is
               justified at least during a certain period of time.
\end{description}
(The assumption (A1) implies that the two spaces
 possess the identical global topology.)

 Now, let us consider two nearby spaces ${\cal G}$ and ${\cal G}'$ in
 ${\cal S}_N$;  their discrepancy is represented by a small difference in
 their spatial metrics,
\begin{equation}
\g_{ab}:=h'_{ab}-h_{ab} \ \ ,
\label{eq:gamma}
\end{equation}
where $h_{ab}$ and $h'_{ab}$ are the spatial metrics for ${\cal G}$ and
${\cal G}'$, respectively.
 We treat $\g_{ab}$ as a  small quantity and
indicate its order of magnitude by $O(\g)$.

By means of the variation formulas for the spectra~\cite{MS-evolution},
we get
\begin{equation}
\d \ln \l_n :=\frac{\l_n'-\l_n}{\l_n}
= -\<\overline{\g}_{ab}{\>_{}}_{n}
                -\frac{1}{2}\<\g {\>_{}}_{n} \ \ ,
\label{eq:dloglambda-distance}
\end{equation}
where  $\g:=h^{ab}\g_{ab}$ and
 $\overline{\g}_{ab}:=\g_{ab}-\frac{1}{2}\g h_{ab}$.

Since Eq.(\ref{eq:dloglambda-distance}) plays the role of  our key equation,
let us briefly review how to derive it.
The discrepancy of metrics, $\g_{ab}:=h'_{ab}-h_{ab}$, causes
the variation of the Laplacian, $\d \Delta$, which
induces the discrepancy of the $n$-th spectrum, $\d \l_n:=\l_n'-\l_n$.
First we recall there is a well-known relation (``Fermi's golden rule")
\begin{equation}
\d \l_n = - \<\d \Delta {\>_{}}_{n}:=
-\int_\Sigma f_n \d \Delta f_n \/ \ \ .
\label{eq:golden}
\end{equation}
 Thus we
need the expression for $\d \Delta$. For this purpose we take the
variation of both sides of
$\Delta f= \/^{-1}\partial_a(\ h^{ab}\partial_b f \/ )$ with respect to
$h_{ab}$, yielding
\[
\d \Delta f = \frac{1}{2}\partial_a \g \partial^a f
-\frac{1}{\/} \partial_a ({\g^a}_c\partial^c f\/) \ \ .
\]
Thus we get,
\[
\<\d \Delta{\>_{}}_{n}
       = \frac{1}{2}\int f_n \partial_a \g  \partial^a f_n \/
            - \int f_n \partial_a ({\g^a}_c\partial^c f_n \/)\ \ .
\]
By suitable partial integrals, this expression along with
Eq.(\ref{eq:golden}) yields Eq.(\ref{eq:dloglambda-distance}).

Let us now consider the general form,  Eq.(\ref{eq:d_N_general}).
We insert $\l'_n=\l_n + \d \l_n$ into
the R.H.S. (right-hand side) and expand it in terms of $\d \l_n$.
With the help of Eq.(\ref{eq:dloglambda-distance}), then,
the leading term in Eq.(\ref{eq:d_N_general}) becomes~\cite{MS-evolution}
\begin{equation}
d_N ({\cal G},{\cal G}')
= \frac{1}{2}{\cal F}''(1)  \sum_{n=1}^{N}
    \left(\<\overline{\g}_{ab}{\>_{}}_{n} +
              \frac{1}{2}\<\g {\>_{}}_{n}\right)^2 \ \ .
\label{eq:d_N-close}
\end{equation}
We thus observe a prominent feature; {\it
whenever $\cal G$ and ${\cal G}'$ are very close to each other
in ${\cal S}_N$,   the leading term  of the spectral distance
$d_N ({\cal G},{\cal G}')$  is universally given by
Eq.(\ref{eq:d_N-close}), irrespective either of the detailed form of
the spectral distance or of the gravity theory.}

We note that  the choice of the function ${\cal F}(x)$  affects
only the unimportant numerical coefficient of
$d_N ({\cal G},{\cal G}')$ in the leading order,
irrespective of the cut-off order $N$.
Thus, it suffices to consider a particular form given by Eq.(\ref{eq:d_N}).
Then  ${\cal F}''(1)=\frac{1}{8}$, so that
 Eq.(\ref{eq:d_N-close}) becomes
\begin{equation}
d_N ({\cal G},{\cal G}')
= \frac{1}{16}  \sum_{n=1}^{N}
    \left(\<\overline{\g}_{ab}{\>_{}}_{n} +
              \frac{1}{2}\<\g {\>_{}}_{n}\right)^2
 =\frac{1}{16}
        \vec{\mbox{\boldmath $\g$}}
                     \cdot\vec{\mbox{\boldmath $\g$}} \ \ ,
\label{eq:d_N-close3}
\end{equation}
where in the last line, $\vec{\mbox{\boldmath $\g$}}$ indicates a vector in
$\mbox{\bf R}^N$ whose $n$-th component is
$
\g_n:=\<\overline{\g}_{ab}{\>_{}}_{n} + \frac{1}{2}\<\g {\>_{}}_{n}
$.
We can further derive the expressions for $\dot{d}_N ({\cal G},{\cal G}')$ and
$\ddot{d}_N ({\cal G},{\cal G}')$ (For more details, see
Ref.\cite{MS-evolution}).

Equation Eq.(\ref{eq:d_N-close3}) plays the role of the central
equation for our investigation on the evolution of the discrepancy
between $\cal G$ and ${\cal G}'$.

\section{Quantitative description of
the relation between two nearby geometries}
\label{section:III}
\subsection{Models to be investigated}
\label{subsection:III-1}

Let us now investigate explicitly the discrepancy of
two specific nearby geometries.\footnote{
We set $D=4$ hereafter unless otherwise stated.
}

As a ``model" universe,
 we construct a spacetime in the form of
$T^3 \times \mbox{\bf R}$  as follows\footnote{In this paper,
Roman indices (such as
$a$, $b$, $\cdots$, $k$, $l$, $\cdots$) indicate the spatial
indices and they are raised and lowered by the spatial metric.
On the other hand, Greek indices
(such as $\a$, $\b$, $\cdots$) indicate the spacetime indices and
they are raised and lowered by the spacetime metric.
}:
Let $x^a$ ($a=1,2,3$) be the Cartesian  spatial
coordinates on ${\mbox{\bf R}}^3$.
By imposing the identification $x^a \sim x^a +1$ ($a=1,2,3$), then,
$x^a$ ($a=1,2,3$) turns to a coordinate on a circle $S^1$,
taking  its value in $[0,1]$ with the identification $0 \sim 1$.
We then  consider a metric defined by
\begin{equation}
ds^2=-dt^2 + a(t)^2 \d_{ab} dx^a dx^b \ \ ,
\label{eq:metric-model}
\end{equation}
where $a(t)$ is the scale-factor of the model.
The $t=constant$ spatial section
of this model, $\Sigma_t$, corresponds to $\cal G$ in the previous section.

As for  the ``real Universe",
a perturbed geometry around the above model is chosen:
We prepare a spacetime $T^3 \times \mbox{\bf R}$  with
the same spatial coordinates as above,
${x'}^a \in [0,1]$ with $0 \sim 1$.
The metric is then given by
\begin{equation}
ds^2=-dt^2 + h_{ab}d{x'}^a d{x'}^b \ \ ,
\label{eq:spacetime-metric}
\end{equation}
where the spatial metric $h_{ab}$ is given by
\begin{eqnarray}
h_{ab}(t, {x'}^a)&=&a^2(t)\  \th_{ab}(t, {x'}^a)
      :=a^2(t) \left( \d_{ab} + {\tg}_{ab}(t, {x'}^a)\right) \ \ , \nonumber \\
{\g}_{ab}(t, {x'}^a):&=& a^2(t)\  \tg_{ab}(t, {x'}^a) \ \ .
\label{eq:spatial-metric}
\end{eqnarray}
The $t=constant$ spatial section
of this spacetime, $\Sigma'_t$, corresponds to ${\cal G}'$
in the previous section.

For later convenience, let $T^3(1)$ denote the regular 3-torus that is
conformally equivalent to $\cal G$;
$T^3(1):=[0,1]\times[0,1]\times[0,1]/_\sim$, with the standard metric
$dl^2=\d_{ab}dx^a dx^b$ ($/_\sim$ indicates
the  identification for making the regular 3-torus).
Let  $\{ \tilde{\l}_n \}_{n=0,1,2,\cdots}$ and
$\{ \tf_n \}_{n=0,1,2,\cdots}$ are, respectively,
 the spectra and the eigenfunctions of
 the Laplacian on $T^3(1)$, $\tilde{\Delta}:=\d^{ab}\partial_a \partial_b$.

\subsection{Spectral representation of geometrical quantities}
\label{subsection:III-2}

In view of Eqs.(\ref{eq:scalar-perturbation}),
(\ref{eq:vector-perturbation}) and
(\ref{eq:tensor-perturbation}),
the metric perturbation $\tg_{ab}$ in Eq.(\ref{eq:spatial-metric})
can be expanded in terms of
$\tf_A  \th_{ab}$, $\ts^A_{ab}$, $\tilde{\zeta}^A_{ab}$ and
$\tw^A_{ab}$:
\begin{equation}
\tilde{\g}_{ab} = \frac{1}{3}\sum_A \gc_{\ A}\ \tf_A \ \th_{ab}
                    + {\sum_A}' \gs_{\ A}\  \ts^A_{ab}
                    + {\sum_A}'  \gv_{\ A}\  \tilde{\zeta}^A_{ab}
                    +\sum_A \gt_{\ A}\  \tw^A_{ab} \ \ .
\label{eq:metric-expansion}
\end{equation}
Here the prime symbol in ${\sum_A}' $ implies omitting  the zero-mode
of $\tilde{\Delta}$ from the summation; the coefficients $\gc_{\ A}$,
$\gs_{\ A}$, $\gv_{\ A}$ and $\gt_{\ A}$ are attributed to the perturbations of
 conformal-type, scalar-type, vector-type and tensor-type, respectively.
We consider any perturbed quantity  up to $O(\g)$ hereafter.
From Eq.(\ref{eq:metric-expansion}), we get
\begin{equation}
\g= h^{ab}\ \g_{ab} = \sum_A \gc_{\ A}\ \tf_A \ \ .
\label{eq:trace-gamma-expansion}
\end{equation}

Now from Eqs.(\ref{eq:spatial-metric}) and (\ref{eq:metric-expansion}),
it is straightforward to compute
$\mbox{\boldmath $R$}_{ab}$, $\mbox{\boldmath $R$}$ and
$r_{ab}$ (see Eq.(\ref{eq:e/r})) for ${\cal G}'$.
Extracting the $f_A$-components
of $\mbox{\boldmath $R$}$ and $D^a D^b r_{ab}$
(according to Eq.(\ref{eq:fourier-component})), thus, we get
\begin{eqnarray}
\mbox{\boldmath $R$}_A &=& \frac{2}{3} \l_A (\gc_{\ A}+\gs_{\ A}) \sqrt{a^3}
\ \ (A \neq 0) ,       \nonumber \\
(D^a D^b r_{ab})_A &=& -\frac{1}{9} \l_A (\gc_{\ A}+\gs_{\ A}) \sqrt{a^3}
 \ \ (A \neq 0) .
\label{eq:R_A}
\end{eqnarray}
The two formulas in (\ref{eq:R_A}) are equivalent
to each other due to the Bianchi identity,
$D^a(\mbox{\boldmath $R$}_{ab}-\frac{1}{2}\mbox{\boldmath $R$} h_{ab})
=0$.\footnote{
For, the spectral representation of the Bianchi identity becomes
$\mbox{\boldmath $R$}_A + \frac{6}{\l_A}(D^aD^b r_{ab})_A=0$
($A \neq 0$)~\cite{MS-evolution}.
}

In the same manner, $K$ and $\e_{ab}$ (see Eq.(\ref{eq:e/r})) are
expanded  as
\begin{eqnarray}
K&=&K_{\rm av}(1+ {\sum_A}'\k_A \tf_A )\ \ , \nonumber \\
\e_{ab}&=& K_{\rm av}\left\{ {\sum_A}'(a^2 \bs_A) \ \ts^A_{ab}
                  + {\sum_A}'(a^2 \bv_A) \ \tilde{\zeta}^A_{ab}
                  + \sum_A (a^2 \bt_A) \ \tw^A_{ab} \right\} \ \ .
\label{eq:K/e-expansion}
\end{eqnarray}

Extracting the $f_A$-components of
$K$ and $D^a D^b \e_{ab}$,
 thus, we get
\begin{eqnarray}
K_A &=&
\left\{
       \begin{array}{rl}
           3 H \sqrt{V} & \  (A=0)  \ \ ,  \\
           3 H \k_A  \sqrt{a^3} & \  (A \neq 0)  \ \ ,   \\
       \end{array}
\right.
                    \nonumber \\
(D^a D^b \e_{ab})_A &=&
    2 H \l_A \bs_A  \sqrt{a^3} \ \ (A \neq 0) \ \ .
\label{eq:K_A}
\end{eqnarray}
Here we used the relation $K_{\rm av}=3H:=3\dot{a}/a$;
To obtain the last equation, the second equation in (\ref{eq:sigma})
has been used.

\subsection{Spectral representation of matter quantities}
\label{subsection:III-3}

We consider
the perfect-fluid type matter,
$T^{\a \b}=(\rho + p)u^\a u^\b + p g^{\a \b}$,  where
$u^\a$ is the 4-velocity of the fluid element ($u^\a u_\a=-1$).
The matter velocity, $u^a$ $(a=1,2,3)$, is regarded as $O(\g)$.
 The standard parameters are introduced for later use:
$\nu:=p/\rho$ ($0 \leq \nu \leq 1/3$ for normal matter)
and $c_s:=\sqrt{dp/d\rho}$, the sound velocity of the matter
relative to the fluid element.

We  expand  $\rho$ and $p$ as
\begin{equation}
\rho = \rho_{\rm av}(1+ \d ), \ \
p = p_{\rm av}+ \rho_{\rm av} c_s^2 \d   \ \ , \ \
     \d  =  {\sum_A}'\d_A \tf_A   \ \ .
\label{eq:rho/p-expansion}
\end{equation}
Then, the spectral representation of $\rho$ becomes
\begin{equation}
  \rho_A =
   \left\{
     \begin{array}{rl}
       \rho_{\rm av} \sqrt{V} & \  (A=0)  \ \ ,   \\
       \d_A \rho_{\rm av} \sqrt{a^3} & \  (A \neq 0)  \ \ . \\
     \end{array}
   \right.
\label{eq:rho_A}
\end{equation}
The expansion of matter $\theta:=D_a u^a$ is  also  expanded as
$\theta = 3H{\sum_A}'\Theta_A \tf_A $,
then its spectral representation becomes
\begin{equation}
\theta_A = 3H\Theta_A \sqrt{a^3}
\ \ (A \neq 0) \ \ .
\label{eq:theta_A}
\end{equation}
Then the divergence of the momentum density of matter,
$J_a =(\rho_{av} + p_{av})u_a + O(\g^2)$,  reduces to
the spectral representation,
\begin{equation}
(\vec{D}\cdot \vec{J})_A \simeq
(\rho_{\rm av} + p_{\rm av}) \theta
    = 3(1+\nu) H \rho_{\rm av}  \Theta_A \sqrt{a^3}
\ \ (A \neq 0) \ \ .
\label{eq:DJ_A}
\end{equation}

\subsection{Spectral representation of constraint equations}
\label{subsection:III-4}

The Hamiltonian constraint is represented
in the spectral form as~\cite{MS-evolution}
\begin{equation}
\mbox{\boldmath $R$}_A
    + \frac{2}{3}\sum_{A'A''}K_{A'} K_{A''}(A'\ A'' \  A) -
    \frac{1}{\a}\rho_A -2\Lambda \sqrt{V}\d_{A0} -
                                 (\e_{ab} \e^{ab})_A=0 \ \ ,
\label{eq:Hamiltonian-constraint_A}
\end{equation}
where $\a:=\frac{c^3}{16 \pi G}$.
(See Eq.(\ref{eq:(lmn)}) for the definition of $(A'\ A'' \  A)$.)
We neglect the quantities of  order $O(\g^2)$.
Then, for $A=0$, Eq.(\ref{eq:Hamiltonian-constraint_A}) becomes
\begin{equation}
6H^2 - \frac{1}{\a}\rho_{\rm av} - 2 \Lambda = 0 \ \ ,
\label{eq:Ha}
\end{equation}
with the help of Eqs.(\ref{eq:K_A}) and (\ref{eq:rho_A}).
 (Here we note $(A'\ A'' \  0)=\frac{1}{\sqrt{V}}\d_{A'A''}$.)
Needless to say, Eq.(\ref{eq:Ha}) coincides with the standard equation
for the (locally) flat Friedmann-Robertson-Walker model.
 On the other hand,  Eq.(\ref{eq:Hamiltonian-constraint_A})
 becomes, for $A \neq 0$,
\begin{equation}
\frac{2}{3}\l_A (\gc_A + \gs_A) + 12 H^2 \k_A
     - \frac{1}{\a}\rho_{\rm av} \d_A =0 \ \ (A \neq 0)\ \  ,
\label{eq:Hb}
\end{equation}
with the help of Eqs.(\ref{eq:R_A}), (\ref{eq:K_A}) and (\ref{eq:rho_A}).
(Here we note
$\sum_{A'A''}K_{A'} K_{A''}(A'\ A'' \  A)
= 18 H^2 \k_A \sqrt{a^3} + O(\g^2)$ for $A \neq 0$.)
With the help of Eq.(\ref{eq:Ha}), Eq.(\ref{eq:Hb}) is represented
as
\begin{equation}
\gc_A + \gs_A =  \frac{9H^2}{\l_A}
 \left\{ (1-\frac{\Lambda}{3H^2})\d_A - 2 \k_A \right\}  \ \ (A \neq 0)\ \  .
\label{eq:Hb'}
\end{equation}

On the other hand,  the momentum constraint in the
spectral form becomes~\cite{MS-evolution}
\begin{equation}
K_A
  + \frac{3}{2}\frac{1}{\l_A} \left\{
              (D^a D^b \e_{ab})_A + \frac{1}{2\a}(\vec{D}\cdot \vec{J})_A
                                   \right\} =0 \ \ (A\neq 0)\ \  .
\label{eq:momentum-constraint_A}
\end{equation}
 With the help of Eqs.(\ref{eq:K_A}) and  (\ref{eq:DJ_A}), thus,
 Eq.(\ref{eq:momentum-constraint_A}) gives
\begin{equation}
(\k_A + \bs_A)+\frac{3}{2}\frac{1}{\l_A}(1+\nu)(3H^2-\Lambda)
\Theta_A=0 \ \ (A \neq 0)\ \ ,
\label{eq:momentum-constraint_A2}
\end{equation}
where Eq.(\ref{eq:Ha}) was used to eliminate $\rho_{\rm av}$.

The coefficients $\k_A$'s and  $\b_A$'s are related to $\dot{\g}_A$'s:
From Eqs. (\ref{eq:spacetime-metric}) and  (\ref{eq:spatial-metric}),
we get
$K_{ab}=\frac{1}{2}\dot{h}_{ab}=$$\frac{\dot{a}}{a}\ h_{ab}
+ \frac{a^2}{2}\dot{\tg}_{ab}$. Together with
Eqs. (\ref{eq:metric-expansion}) and  (\ref{eq:trace-gamma-expansion}),
this relation yields the spectral expressions for
$K$ and $\e_{ab}$ up to $O(\g)$.
Comparing them  with
Eq.(\ref{eq:K/e-expansion}), we finally get
\begin{eqnarray}
\gcdot_0&=& 6(H-\frac{\dot{a}}{a})=:6 \d H \ \ , \ \
\k_A = \frac{1}{6H} \gcdot_A \ \ (A \neq 0) \ \ , \nonumber \\
\bs_A &=& \frac{1}{6H} \gsdot_A \ \ (A \neq 0) \ \ , \ \
\bv_A = \frac{1}{6H} \gvdot_A \ \ (A \neq 0) \ \ , \ \
\bt_A = \frac{1}{6H} \gtdot_A .
\label{eq:beta-gamma}
\end{eqnarray}

Then Eq.(\ref{eq:momentum-constraint_A2}) reduces to
\begin{equation}
(\gc_A + \gs_A \dot{)\ }+\frac{9H}{\l_A}(1+\nu)(3H^2-\Lambda)
\Theta_A=0 \ \ (A \neq 0)\ \ .
\label{eq:momentum-constraint_A3}
\end{equation}
In particular, we note that $\gc_A + \gs_A=constant$ when $\theta=0$.

We have obtained necessary relations to investigate
$\mbox{\boldmath $\g$}_A:=\<\overline{\gamma}_{ab}{\>_{}}_{A}
          +\frac{1}{2}\<\gamma {\>_{}}_{A}$, i.e.
  the $A$-th component of $\vec{\mbox{\boldmath $\g$}}$,
  which is important in view of
  Eq.(\ref{eq:d_N-close3}).
 With the help of Eqs.(\ref{eq:metric-expansion}),
 (\ref{eq:trace-gamma-expansion}),
 (\ref{eq:c}), (\ref{eq:sigma}), (\ref{eq:<zeta>}) and
 (\ref{eq:c^T1}), we get
 after some manipulations,
\begin{equation}
\mbox{\boldmath $\g$}_A =
 \frac{1}{3}\gc_0
+ \frac{1}{3}{\sum_{A'}}'(c^{(1)}_{AA'}+c^{(2)}_{AA'})(\gc_{A'}+\gs_{A'})
+\sum_{A'} c^{(T)}_{AA'} \gt_{A'} \ \ ,
\label{eq:gamma_A}
\end{equation}
where $c^{(1)}_{AA'}$, $c^{(2)}_{AA'}$ and
$c^{(T)}_{AA'}$ are given in Eqs.(\ref{eq:c}) and (\ref{eq:c^T1}).
We note that the vector components do not  appear in Eq.(\ref{eq:gamma_A});
it  can be traced back to the spatial-diffeomorphism invariance of
the spectral distance (Eq.(\ref{eq:d_N_general})).
Thus we can safely omit the vector quantities
(those with suffix $(V)$) hereafter.

\section{Combination of the spectral scheme and
 the linear structure-formation theory}
\label{section:IV}

Looking at  Eq.(\ref{eq:gamma_A}), the behavior of coefficients
$\gc_A$, $\gs_A$ and $\gt_A$ determine the evolution of the spectral
distance $d_N({\cal G}, {\cal G}')$. Considering
Eq.(\ref{eq:metric-expansion}), these behaviors are translated from
the time evolution of $\tg_{ab}$, which can be estimated by the
structure-formation theory.
(The basic results of  the standard linear
structure-formation theory is summarized in {\it Appendix} B ~\cite{PJEP}.)
Here two scales control the situation: The physical wave-length of the
perturbation (of geometry and matter), $\l_{\rm phys}$, and the
causal scale at the time $t$, $c_s t$ ($c_s$ is the sound velocity of
matter).

\subsection{Behavior of coefficients $\g_A$, $\d_A$, $\b_A$ and $\k_A$:
The case of $\l_{\rm phys} \gg c_s t$ and $\theta \neq 0$}
\label{subsection:IV-1}

We first consider the case
(I) $\l_{\rm phys} \gg c_s t$ and $\theta \neq 0$.

Since $\frac{\l_{\rm phys}}{c_s} \gg t$ in this case,
a perturbation mode satisfying this condition
does not oscillate in effect  during the cosmological time-scale $t$.
Comparing the expansions of $\g$, $\d$ and $\theta$ in
Eqs.(\ref{eq:trace-gamma-expansion}),
(\ref{eq:rho/p-expansion}) and (\ref{eq:theta_A})
with their behavior in Eq.(\ref{eq:result1}),
we see that
\begin{equation}
\d_A(t) \propto  t^{\frac{9\nu-1}{3(1+\nu)}}\ \ , \ \
 \gc_A(t) = -\frac{2(1+3\nu)}{\nu (9\nu -1)}\  \d_A \ \ ,  \ \
 \Theta_A (t) = \frac{(1-\nu)(6\nu +1)}{6\nu(1+\nu)}\  \d_A\ \ ,
\label{eq:evolve1}
\end{equation}
where we have used $H(t) =\frac{2}{3(1+\nu)} 1/t + O(\g)$
(the standard result derived from Eq.(\ref{eq:Ha}))
in the last equation.

On the other hand, comparing the expansion of $\tg_{kl}$ in
 Eq.(\ref{eq:metric-expansion}) with their behavior
 in Eq.(\ref{eq:result-gamma1}), we get
\begin{equation}
\gs_A(t),\   \gt_A(t) \propto
      \frac{1}{\nu}\left(\frac{a^2}{k^2 t^2} \right) \d_A (t)
      \propto \frac{1}{\nu}\left(\frac{\l_{\rm phys}}{c_s t} \right)^2 \d_A (t)
      \propto \frac{1}{\nu k^2}t^{-\frac{1-\nu}{1+\nu}} \ \ .
\label{eq:metric-evolve1}
\end{equation}
We omit vector quantities since they do not contribute to the
behavior of the spectral distance.
 Since $\frac{\l_{\rm phys}}{c_s t} \gg 1$,  Eqs.(\ref{eq:evolve1}) and
(\ref{eq:metric-evolve1}) indicate
\[
|\gc_A| \ll |\gs_A|,\   |\gt_A|\ \ ,
\]
which is compatible with Eq.(\ref{eq:result-order1}).

 Now from Eq.(\ref{eq:beta-gamma}) we see that
\begin{equation}
\k_A = \frac{1}{6H} \gcdot_A \propto \gc_A
\propto \frac{1}{\nu}\ t^{\frac{9\nu-1}{3(1+\nu)}}\ \ ,
\label{eq:kappa-evolve1}
\end{equation}
 where we have used $|\gcdot_A| \propto |\gc_A|/t$ since $t \ll \tau$
 (see {\it Appendix} \ref{subsection:B-1}) and $H \propto 1/t$.

 In the same manner, from Eq.(\ref{eq:beta-gamma}),
\begin{eqnarray}
\bs_A &\propto & \gs_A
      \propto  \frac{1}{\nu}\left(\frac{\l_{\rm phys}}{c_s t}\right)^2 \d_A
      \propto  \frac{1}{\nu k^2} t^{-\frac{1-\nu}{1+\nu}} \ \ , \nonumber \\
\bt_A &\propto & \gt_A
      \propto \frac{1}{\nu k^2} t^{-\frac{1-\nu}{1+\nu}} \ \ .
\label{eq:beta-evolve1}
\end{eqnarray}
Comparing these coefficients with $\d_A$ we see that
\begin{equation}
(|\d_A| \sim |\gc_A| \sim  |\k_A|)
\ll (|\gs_A| \sim |\bs_A|, \ |\gt_A| \sim |\bt_A|) \ \ .
\label{eq:order2-evolve1}
\end{equation}

\subsection{Behavior of coefficients $\g_A$, $\d_A$, $\b_A$ and $\k_A$:
The case of $\l_{\rm phys} \gg c_s t$ and $\theta = 0$}
\label{subsection:IV-2}

We next consider the case
(I') $\l_{\rm phys} \gg c_s t$ and $\theta = 0$.

Comparing the expansions of $\g$ and $\d$ in
Eqs.(\ref{eq:trace-gamma-expansion}) and
(\ref{eq:rho/p-expansion})
with their behavior in Eq.(\ref{eq:result2}),
we see that
\begin{equation}
\d_A(t) \propto  t^{\frac{2(1+3\nu)}{3(1+\nu)}}\ \ ,  \ \
 \gc_A(t) = -\frac{2}{1+\nu} \d_A \ \ .
\label{eq:evolve2}
\end{equation}

Next, comparing the expansion of $\tg_{kl}$ in
 Eq.(\ref{eq:metric-expansion}) with their behavior
 in Eq.(\ref{eq:result-gamma2}), we get
\begin{equation}
\gs_A(t), \ \gt_A(t) \propto
     \left(\frac{a^2}{k^2 t^2} \right) \d_A (t)
     \propto \left(\frac{\l_{\rm phys}}{c_s t} \right)^2 \d_A (t)
      \propto \frac{1}{k^2}t^0  \ \ .
\label{eq:metric-evolve2}
\end{equation}

From Eqs.(\ref{eq:evolve2}) and (\ref{eq:metric-evolve2}), we observe that
\[
|\gc_A| \ll |\gs_A|,  \ |\gt_A| \ \ ,
\]
which is compatible with Eq.(\ref{eq:result-order2})

 In the same manner as in \S \ref{subsection:IV-1},
we get  from Eq.(\ref{eq:beta-gamma}) that
\begin{eqnarray}
\k_A & = & \frac{1}{6H} \gcdot_A
          \propto \gc_A \propto t^{\frac{2(1+3\nu)}{3(1+\nu)}}
                                                \ \ , \nonumber \\
\bs_A & = &   \frac{1}{6H} \gsdot_A
       \propto   \nu \d_A
    \propto \nu t^{\frac{2(1+3\nu)}{3(1+\nu)}} \ \ , \ \
\bt_A =  \frac{1}{6H} \gtdot_A
        \propto \nu t^{\frac{2(1+3\nu)}{3(1+\nu)}} \ \ ,
\label{eq:beta-evolve2}
\end{eqnarray}
where we note $H^{-1}\gsdot_A,  H^{-1}\gtdot_A \propto \nu \d_A$ due to
Eq.(\ref{eq:result-gamma-dot2}).
Comparing Eqs.(\ref{eq:evolve2})-(\ref{eq:beta-evolve2})
with each other,
we thus get
\begin{eqnarray}
&& (|\d_A| \sim |\gc_A|) \sim  (|\k_A| \sim |t\gcdot_A|)
   \sim (|\bs_A| \sim t|\gsdot_A|), \  (|\bt_A| \sim t|\gtdot_A|) \nonumber \\
&& \qquad \qquad  \ll |\gs_A|, |\gt_A| \ \ .
\label{eq:order2-evolve2}
\end{eqnarray}

\subsection{Behavior of coefficients $\g_A$, $\d_A$, $\b_A$ and $\k_A$:
 The case of $\l_{\rm phys} \ll c_s t$}
\label{subsection:IV-3}

Finally, we consider the case
(II) $\l_{\rm phys} \ll c_s t$.
A perturbation mode satisfying this condition oscillates many times
within the cosmological time-scale.

Comparing the expansions of $\g$, $\d$ and $\theta$ in
Eqs.(\ref{eq:trace-gamma-expansion}),
(\ref{eq:rho/p-expansion}) and (\ref{eq:theta_A})
with their behavior in Eq.(\ref{eq:result3}),
we observe that
\begin{eqnarray}
\d_A(t) &\propto & \frac{1}{\sqrt{k}} t^{-\frac{1-3\nu}{3(1+\nu)}} \cdot Osc
    \ \ ,  \ \
 \gc_A(t)  \simeq   -\d_A (t) \ \ , \nonumber \\
 \Theta_A (t) & \propto &
        \frac{t}{\tau} \ \frac{1}{\sqrt{k}}  t^{-\frac{1-3\nu}{3(1+\nu)}}
         \cdot Osc'  \ \
            \simeq  \sqrt{k} t^{\frac{2\nu}{1+\nu}} \cdot Osc'\ \ .
\label{eq:evolve3}
\end{eqnarray}
Here $Osc$ and $Osc'$ indicate the oscillatory factors with the period
$\tau = 2\pi a/k c_s$ whose relative phase-difference is $\pi/2$;
we have used $H \propto 1/t$ and assumed $c_s \approx constant$ to reach
the final estimation in the third relation.

We now compare
 Eqs.(\ref{eq:beta-gamma}) and (\ref{eq:metric-expansion})
 with Eq.(\ref{eq:result-order3}).
 For instance, we can estimate as $|\bs_A| \propto |H^{-1}\gsdot_A| \propto
 \frac{t}{\tau}|\tau \gsdot_A|$$\sim
 \left( \frac{\l_{\rm phys}}{c_s t} \right) |\d_A| \ll |\d_A| $.
 In this manner,  we get
\begin{equation}
(|\d_A| \sim |\gc_A|) \sim  (|\gs_A|,  \ |\gt_A|)
\gg (|\k_A| \sim |\bs_A|,  \  |\bt_A|)    \ \ .
\label{eq:order-evolve3}
\end{equation}

\section{Evolution of the discrepancy between
two nearby geometries}
\label{section:V}
\subsection{Some useful propositions}
\label{subsection:V-1}

Let us  prepare propositions that become useful soon.

 We first introduce some  notations. Let
$\vec{V}=(V_1, V_2, V_3)$ and $\vec{W}=(W_1, W_2, W_3) $ be vectors,
whose components are non-negative integers, viz.
 $\vec{V}, \vec{W} \in {\mbox{\bf N}_0}^3$
(where $\mbox{\bf N}_0 :=\{0 \}\cup \mbox{\bf N}$).

Now, $\# (\vec{V})$ is defined as
the number of zero elements in $\vec{V}$.\footnote{
For example, $\#(1,2,3)=0$, $\#(3,0,2)=1$ and $\#(0,0,0)=3$.}

Next, we say that $\vec{V}$ and $\vec{W}$ are  {\it compatible} with each other
 and write $\vec{V} \sim_{comp} \vec{W} $
 when $\# (\vec{V}) = \# (\vec{W})$ and the positions of their zero
elements are identical.\footnote{
For example, $(1,3,0) \sim_{comp}(2,1,0)$, $(2,0,3)\sim_{comp}(3,0,3)$;
$(2,3,0) \not\sim_{comp}(0,1,2)$.}

 Then we define $\vec{V}_{\|(\vec{W})}$ to be
 the component of $\vec{V}$ that is
compatible with $\vec{W}$, viz.
$\vec{V}_{\|(\vec{W})}:=
\Bigl((1-\d_{W_1,0})V_1,\ (1-\d_{W_2,0})V_2,\
(1-\d_{W_3,0})V_3 \Bigr)$.\footnote{
For example, when $\vec{V}=(2,3,1)$ and $\vec{W}=(0,1,2)$,
$\vec{V}_{\|(\vec{W})}=(0,3,1)$.}

Similarly, we define $\vec{V}_{\perp (\vec{W})}$ to be
  the component of $\vec{V}$ that is causing
the  incompatibility  with $\vec{W}$, viz.
$\vec{V}_{\|(\vec{W})}:=\vec{V}-\vec{V}_{\|(\vec{W})}
=\left(\d_{W_1,0} V_1, \d_{W_2,0} V_2, \d_{W_3,0} V_3 \right)$.

We may simply write $\vec{V}_\|$ and $\vec{V}_\perp$
instead of $\vec{V}_{\| (\vec{W})}$ and $\vec{V}_{\perp (\vec{W})}$,
respectively, whenever obvious.

As is explained in {\it Appendix} A, the basic modes of $\Delta$ on
the model space $\Sigma$ are labeled by the pair of two vectors,
$(\vec{n}, \vec{\s})$,
($n_1,\ n_2,\  n_3 \in \mbox{\bf N}_0$; $\s_1$, $\s_2$, $\s_3 \in \{ 0,1 \}$).
We often write simply $A$ for $(\vec{n}, \vec{\s})$.
Note that $A=(\vec{0}, \vec{\s})$ automatically implies
$A=(\vec{0}, \vec{0})$ since $f_{\vec{n}=\vec{0}}^{\ (\s=\vec{0})}=\frac{1}{\sqrt{a^3}}$
is the only possible eigenfunction for $\vec{n}=\vec{0}$
(see Eq.(\ref{eq:f_A})).

For $A=(\vec{n},\vec{\s})$ and $A'=(\vec{n}',\vec{\s}')$, the quantity
 $\# \left(\vec{\s}_{\|(\vec{n}')} \right)$ (the number of zero
 components in  $\vec{\s}_{\|(\vec{n}')}$) has just been defined above.

Now we can state the proposition.  Let $c_{AA'}$ be
any function of $A=(\vec{n},\vec{\s})$ and $A'=(\vec{n}',\vec{\s}')$
containing  the factor $(-)^{\# \left(\vec{\s}_{\|(\vec{n}')} \right)}$.
 Then, for $\a_{A}$ which is  any function of $A$,
it follows that
\begin{description}
\item{\it Proposition  1}
\[
 {\sum_{A}}' c_{AA'}\a_{A'} =
   \left\{
     \begin{array}{rl}
      0 & \  (\vec{n}' \neq \vec{0})  \ \ ,   \\
      \a_0 {\sum_A}'c_{A0}  & \  (\vec{n}' = \vec{0})  \ \ . \\
     \end{array}
   \right.
\]
(Here $\a_0$ and $c_{A0}$ are abbreviations for
$\a_{(\vec{0},\vec{0})}$ and $c_{A\ (\vec{0},\vec{0})}$, respectively;
the prime symbol in ${\sum_{A}}'$ indicates $A=(\vec{0},\vec{0})$ is
not included in the summation.) For the proof, we note that
the L.H.S. (left-hand side) contains a factor
$\sum_{\vec{\s}} (-)^{\# \left( \vec{\s}_{\|(\vec{n}')} \right)}$.
This factor is nothing but the summation of $+1$'s  and
the same number of $-1$'s (so it vanishes), unless
  ``$\vec{\s}_{\|(\vec{n}')} \equiv \vec{0}$ independently of
  $\vec{\s}$"~\footnote{
For illustration, suppose $\vec{n}'=(1,0,0)$. Then,
$\vec{\s}_{\|(\vec{n}')}=(\s_1,0,0)$ so that
$\# \left( \vec{\s}_{\|(\vec{n}')} \right)=3,2$ for $\s_1=0,1$,
respectively; then
$(-)^{\# \left( \vec{\s}_{\|(\vec{n}')} \right)}=-1,1$ for
$\s_1=0,1$, respectively. Thus
$\sum_{\vec{\s}} (-)^{\# \left(  \vec{\s}_{\|(\vec{n}')} \right)}=0$.
}
; it means  that the L.H.S. vanishes unless $\vec{n}'=\vec{0}$, so that
 the proposition follows.
\end{description}
As a corollary, it immediately follows that
\begin{description}
\item{{\it Corollary of Proposition  1}}
\[
 {\sum_{A}}' {\sum_{A'}}' c_{AA'}\a_{A'} = 0\ \ .
\]
\end{description}
Now, we show another proposition. Let
$A=(\vec{n},\vec{\s})$ and $A'=(\vec{n}',\vec{\s}')$ once more.
It may be clear now what $\vec{n}_{\|(\vec{n}')}$ indicates, and so does
$\d_{\vec{n}',2\vec{n}_{\|(\vec{n}')}}$.
Then this time, let $c_{AA'}$ and  ${c'}_{AA'}$ be any functions which contain
the factor
$(-)^{\# \left( \vec{\s}_{\|(\vec{n}')} \right)}
\d_{\vec{n}',2\vec{n}_{\|(\vec{n}')}} \d_{\vec{\s}',\vec{0}}$.
 Then, for $\a_{A'}$ and $\b_{A'}$ that are  any functions of $A'$, it follows
\begin{description}
\item{\it Proposition  2}
\[
{\sum_{A}}' \left({\sum_{A'}}' c_{AA'}\a_{A'}\right)
         \left({\sum_{A''}}'c'_{AA''}\b_{A''}\right)
      ={\sum_{A}}' {\sum_{A'}}' c_{AA'}c'_{AA'} \a_{A'}\b_{A'}
\]
We note that the L.H.S. contains a factor
$u:=\sum_{\s}(-)^{\# \left( \vec{\s}_\|(\vec{n}')\right) }$
$(-)^{\# \left(\vec{\s}_\|(\vec{n}'') \right)}$.
Every term in $u$
 is $1$ when $\vec{n}' \sim_{comp} \vec{n}''$ so that
 $u$ becomes
  some (non-zero) natural number when $\vec{n}' \sim_{comp} \vec{n}''$;
  on the other hand,
  when
$\vec{n}' \not\sim_{comp} \vec{n}''$,  $u$ is the summation
of $+1$'s and the same number of $-1$'s so that it vanishes.
It means that the factor  $u$ can be regarded as
$\propto \d_{\vec{n}',  \vec{n}''}$
 because of the presence of
$\d_{\vec{n}',2\vec{n}_{\|(\vec{n}')}}\d_{\vec{n}'',2\vec{n}_{\|(\vec{n}'')}}$
 on the L.H.S.  Thus, noting the presence of
another factor $\d_{\vec{\s}',\vec{0}}\d_{\vec{\s}'',\vec{0}}$ on the L.H.S.,
the claim follows.
\end{description}

Finally, suppose $c_{AA'}$ satisfies  the same condition as
in {\it Proposition 2}. Then
\begin{description}
\item{\it Proposition  3}
\[
 {\sum_{A}}'(\a_0 + {\sum_{A'}}' c_{AA'}\a_{A'})^2
 = \a_0^2 {\sum_{A}}'1 + {\sum_{A}}'{\sum_{A'}}'c_{AA'}^2 \a_{A'}^2 \ \ .
\]
It immediately follows from {\it Proposition 1}, its {\it Corollary} and
{\it Proposition 2}.
\end{description}

\subsection{Estimating the time evolution of $d_N ({\cal G},{\cal G}')$}
\label{subsection:V-2}

Now, let us look at Eqs.(\ref{eq:d_N-close3}) and (\ref{eq:gamma_A}).
With the help of {\it Proposition 2} and {\it Proposition 3},
we can simplify the expression for ${\mbox{\boldmath $\g$}_A}^2$.
We assume that the coefficients
$\gc_A$,  $\gs_A$, and  $\gt_A$ are invariant under the
permutation $\tau$ of $A$, $\tau: A \mapsto \tau (A)$.
(More explicitly, $\tau: (\vec{n}, \vec{\s}) \mapsto
\Bigl(\tau (\vec{n}), \tau (\vec{\s}) \Bigr)$.)
Then, we note that any term in $\vec{\g} \cdot \vec{\g}$ of the form
\[
{\sum_{A}}' {\sum_{A'}}' c_{AA'} c^{(T)}_{AA'} \g_{A'} \g^{(T)}_{A'}
\]
vanishes unless $c_{AA'}=c^{(T)}_{AA'}$ because
$c^{(T)}_{AA'}$ is odd for the permutation $\s: AA' \mapsto \s(A)\s(A')$
(see Eq.(\ref{eq:c^T2})),
while other coefficients $c^{(1)}$,  $c^{(2)}$ and $c^{(3)}$ as well as
$\gc_A$,  $\gs_A$, and  $\gt_A$  are
even under the same permutation (see Eq.(\ref{eq:c})).
In a similar argument of permutations,
the contribution of $c^{(T)}_{A0}$ to ${\mbox{\boldmath $\g$}_A}^2$
is only in the form of ${{\sum}'_{A}} {c^{(T)}_{A0}}^2 {\gt_0}^2 $.

Thus, we get
\begin{eqnarray}
d_N ({\cal G},{\cal G}')
 &=&\frac{1}{16}
        \vec{\mbox{\boldmath $\g$}}
                     \cdot\vec{\mbox{\boldmath $\g$}} \ \ , \nonumber \\
&=&
 \frac{1}{144}N {\gc_0}^2
+ \frac{1}{144}
  {\sum_{A}}'{\sum_{A'}}'(c^{(1)}_{AA'}+c^{(2)}_{AA'})^2(\gc_{A'}+\gs_{A'})^2
    \nonumber \\
&+& \frac{1}{16}{\sum_{A}}'\sum_{A'} {c^{(T)}_{AA'}}^2 {\gt_{A'}}^2 \ \ ,
\label{eq:d_N-result}
\end{eqnarray}
Here we note that $N$ is the number of eigenvalues less than
$\l_N$; the contribution of the term
${{\sum}'_{A}} {c^{(T)}_{A0}}^2 {\gt_0}^2 $
is included in the last term on the R.H.S.

We pay attention to  the three typical types of  perturbations:
\begin{description}
\item{(I)}  The case of $\l_{\rm phys} \gg c_s t$ and $\theta \neq 0$
\item{(I')} The case of $\l_{\rm phys} \gg c_s t$ and $\theta = 0$
\item{(II)} The case of $\l_{\rm phys} \ll c_s t$
\end{description}
For the case (I), the behavior of $d_N ({\cal G},{\cal G}')$ is
 controlled by the behavior of
$(\gc_{A'}+\gs_{A'})^2 \simeq {\gs_{A'}}^2$ and  ${\gt_{A'}}^2$, due to
Eq.(\ref{eq:order2-evolve1}). The situation is same for
the case (I'), too,  due to  Eq.(\ref{eq:order2-evolve2}).
On the other hand, for the case (II),
the behavior of $d_N ({\cal G},{\cal G}')$ is determined by
the behavior of all ${\gc_0}^2$,  $(\gc_{A'}+\gs_{A'})^2$ and ${\gt_{A'}}^2$
because of Eq.(\ref{eq:order-evolve3}).

These observations enable us to estimate the time-dependence of ${\g_A}^2$.
First, looking at the explicit expression for $c_{AA'}$'s (Eqs.(\ref{eq:c}) and
(\ref{eq:c^T2})), the summations over  $\s$ and ${\s}'$ on the R.H.S. of
Eq.(\ref{eq:d_N-result}) yield  no essential contributions to
$d_({\cal G}, {\cal G}')$ within the present order of accuracy. Thus,
we can safely replace  $A$ with the coordinate wave-length
$\vec{k}$ defined as $\vec{k}=2\pi \vec{n}$.
Then we may write $\g_k$ instead of $\g_A$, so that $d_N ({\cal G},{\cal G}')$
is estimated as  $d_N ({\cal G},{\cal G}') \sim \int \g_k^2 d\mu(k)$, where
$\mu(k)$ is a suitable measure for $k$.
Then the estimation of ${\g_k}^2$ for the cases (I), (I') and (II) can be
summarized as (see Eqs.(\ref{eq:metric-evolve1}),
(\ref{eq:metric-evolve2}) and (\ref{eq:evolve3}))
\begin{equation}
{\g_k}^2 \simeq
   \left\{
     \begin{array}{rl}
  \frac{D(k)}{\nu^2 k^4}t^{-\frac{2(1-\nu)}{1+\nu}} & \  {\rm for \ (I)}
                                                                \ \ ,   \\
  \frac{D(k)}{k^4} t^{0} & \  {\rm for \ (I')}   \ \ ,   \\
  \frac{D(k)}{k} t^{-\frac{2(1-3\nu)}{3(1+\nu)}}\cdot Osc^2  & \
                          {\rm for \ (II)}   \ \ ,   \\
     \end{array}
   \right.
\label{eq:result}
\end{equation}
where $D(k)$ denotes the extra $k$-dependence coming from
$(a)$ the coefficients $(c^{(1)}_{AA'}+c^{(2)}_{AA'})^2$ and
${c^{(T)}_{AA'}}^2$,
and  $(b)$ the spectral property of the perturbation $\tg_{ab}$, i.e.
the explicit value of
 $\g_A$'s in Eq.(\ref{eq:metric-expansion}).

 The $k$-dependence coming from $(a)$ is  determined without
 ambiguity by the definition of $c_{AA'}$'s (Eqs.(\ref{eq:c}) and
 (\ref{eq:c^T2})). As is explained after Eqs.(\ref{eq:c}) and
 (\ref{eq:c^T2}), the coefficients $c_{AA'}$'s do  not depend on
the magnitude $k$, but
only depend on  the relative direction between $\vec{k}$
and $\vec{k}'$~\footnote{
Note that $\vec{k}=2\pi \vec{n}$ and $k^2= 4\pi^2 \vec{n}\cdot \vec{n}$
(Eq.(\ref{eq:f_A}) and below).
}.
 The summation of both $\vec{k}$ and $\vec{k}'$ ($A$ and $A'$)
 in Eq.(\ref{eq:d_N-result}) ensures the summation for all the
  relative directions $\vec{k}$ and $\vec{k}'$, which yields only
  minor corrections.
The $k$-dependence coming from $(b)$ depends
 on  the initial condition for $\tg_{ab}$.
 As one natural assumption for the initial condition for $\tg_{ab}$,
 we here postulate that the $k$-dependence coming from $(b)$ shows
 no peak at a particular scale but  a mild $k$-dependence.

In any case $D(k)$ is not essential for the $t$-dependence of
$d_N({\cal G}, {\cal G}')$ (see arguments after Eqs.(\ref{eq:final-result3}) and
(\ref{eq:final-result4})).
We here take care of  $D(k)$ by postulating a power-law,
$D(k)\propto k^{-p}$ for (I) and (I'), and $\propto k^{-q}$ for (II).

We can estimate  $d_N ({\cal G},{\cal G}')$
by converting the summation ${\sum_A}'$ in Eq.(\ref{eq:d_N-result})
to $k$-integral from  $k=1$ (because of the $T^3$ structure of the space)
 up to $k=k_N$,  the maximum
value for $k$ determined by the cut-off order $N$.
(It is a standard way of estimating a summation in statistical physics, and
 considering Eq.(\ref{eq:lambda_A}), this is
a reasonable estimation when the universes are sufficiently large.)
We consider two situations,  when we are studying
\begin{description}
 \item{\it (A)} relatively  shorter scale geometry as well as
               the global features of the Universe, and
\item{\it (B)}  only the most global features of the Universe,
\end{description}
where the horizon scale at the time is the natural standard for
$``short"$ or $``long"$.
\begin{description}
\item{\underline{\it Situation (A)}:} This situation
 corresponds to $k_N > k_H$. Here
 $k_H:=2\pi a /c_s t \propto t^{-\frac{1+3\nu}{3(1+\nu)}}$,
  the wave-number corresponding to the horizon scale at $t$.
 From $k=1$ to $k=k_H$ in the integral region,
 we can use the result for the case (I) (or (I'));
from $k=k_H$ to $k= k_N$, we can use the result for the case (II).
Thus we estimate, based on the case (I) for $k \in [1,k_H]$,
\begin{eqnarray}
d_N ({\cal G},{\cal G}') & \simeq &
A' t^{-\frac{2(1-\nu)}{1+\nu}} \int_1^{k_H} \frac{1}{k^{4+p}}k^2 dk
+ B' t^{-\frac{2(1-3\nu)}{3(1+\nu)}} \int_{k_H}^{k_N} \frac{1}{k^{1+q}}
\cdot Osc^2 \  k^2 dk \nonumber \\
& \simeq & A \xi(t)\  t^{-\frac{2(1-\nu)}{1+\nu}}
 + B {k_N}^{2-q} \  \eta (t)\ t^{-\frac{2(1-3\nu)}{3(1+\nu)}}  \ \ .
\label{eq:final-result1}
\end{eqnarray}
Here, $A$ and $B$ are appropriate positive numerical factors that make
the expression dimension-free;
$\xi(t):=1-C t^{\frac{(1+p)(1+3\nu)}{1+\nu}}$
and $\eta(t):=1-C't^{-\frac{(2-q)(1+3\nu)}{3(1+\nu)}}$ with
$C$ and $C'$ being  appropriate positive numerical factors that
make $\xi(t)$ and $\eta(t)$ are positive for $t>\exists t_0$.
We note that $\xi(t) \rightarrow 0$ and $\eta(t) \rightarrow 1$
as $t \rightarrow \infty$. Thus as time goes on,
the second term with the behavior $\sim t^{-\frac{2(1-3\nu)}{3(1+\nu)}}$
becomes dominant.

In the same manner, based on the case (I') for $k \in [1,k_H]$,
  we estimate
\begin{equation}
d_N ({\cal G},{\cal G}')
 \simeq  A \xi(t)
 + B {k_N}^{2-q} \  \eta (t)\ t^{-\frac{2(1-3\nu)}{3(1+\nu)}}  \ \ .
\label{eq:final-result2}
\end{equation}

\item{\underline{\it Situation (B)}:}
This situation corresponds to $k_N$ is less than
  $k_H$.
Thus, for the case (I),
\begin{eqnarray}
d_N ({\cal G},{\cal G}') & \simeq &
A' t^{-\frac{2(1-\nu)}{1+\nu}} \int_1^{k_N} \frac{1}{k^{4+p}}k^2 dk
                                                    \nonumber \\
& \simeq & A (1- {k_N}^{-(1+p)}) \  t^{-\frac{2(1-\nu)}{1+\nu}}  \ \ .
\label{eq:final-result3}
\end{eqnarray}
In the same manner, for the case (I'),
we get
\begin{equation}
d_N ({\cal G},{\cal G}')  \simeq  A' (1- {k_N}^{-(1+p)})\ \ .
\label{eq:final-result4}
\end{equation}

\end{description}

Looking at these results,
Eqs.(\ref{eq:final-result1})-(\ref{eq:final-result4}),
we see that $d_N ({\cal G},{\cal G}')$ decreases with time (with
Eq.(\ref{eq:final-result4}) being a marginal case of constancy).
Typical behavior is
$d_N({\cal G}, {\cal G}') \sim t^{-\frac{2(1-3\nu)}{3(1+\nu)}} $ for
the situation {\it (A)}, and $\sim t^{-\frac{2(1-\nu)}{1+\nu}}$ for
the situation {\it (B)}.  It means that, for the situation {\it (B)},
$d_N(\cal G, {\cal G}')$ decreases more  rapidly than the situation
{\it (A)}. This might be interpreted as that, within the linear regime,
{\it the dynamics of the most global structures of the Universe are
approximated  by those of a model quite well,
compared to the shorter-scale features}.

Significantly,   Eqs.(\ref{eq:final-result1})-(\ref{eq:final-result4})
are also showing the scale-dependence of $d_N(\cal G, {\cal G}')$
as an increasing  function of the cut-off scale $k_N$. It indicates
a reasonable result  that {\it the finer structures are observed,
the more discrepancy between the reality and its model is detected}.
This result is consistent with our observation in the previous paragraph.

Finally,
we realize that $D(k)$ has no significant influence on the $t$-dependence of
$d_N(\cal G, {\cal G}')$; rather $D(k)$ for the integral-region (II)
determines the details on how
$d_N(\cal G, {\cal G}')$ depends on the cut-off scale $N$.

We observed that, in the regime where
 the linear approximation of geometry is valid,
{\it $d_N ({\cal G},{\cal G}')$ does not increase with time, rather
it shows the tendency to decrease}.
This result is favorable to the reliability of
describing the Universe approximately by means of a model,
 at least during some period of time. Needless to say,
 the analysis here is limited to
 just one pair of geometries within the linear-regime.
 The present result calls for more
 investigations along the same line on
 the behavior beyond the liner-regime and  wider class of spacetimes.

\section{Summary and Discussion}
\label{section:VI}

  In this paper, we have studied a fundamental issue rooted in  the
foundation of cosmology: To what extent we can rely on a model to
recognize the Universe.  To analyze this issue, we argued that
the concept of ``closeness" or ``discrepancy" between two geometries
was indispensable, so that we have resorted to the spectral scheme
developed in the previous studies.
Then we have estimated
the discrepancy measured by the spectral distance
between two  spatial geometries, ${\cal G}$ and ${\cal G}'$,
that are very close
to each other, mimicking the relation of the Universe and its model.

We have set  ${\cal G}$ to be
a flat $T^3$ geometry constructed
from the flat Friedmann-Robertson-Walker model obtained by
the point-identification; as for  ${\cal G}'$, we have chosen
the geometry generated  by  perturbing  the  geometry ${\cal G}$.
These ${\cal G}$ and ${\cal G}'$ were regarded as the imitation  of
the relation of the reality and its model.

Then the behavior of the
the spectral distance $d_N ({\cal G},{\cal G}')$ has been investigated
explicitly.
With the help of the linear structure-formation theory,
the behavior of $d_N ({\cal G},{\cal G}')$ has been estimated
as Eqs.(\ref{eq:final-result1})-(\ref{eq:final-result4}).
In particular, it is an important observation that
the geometry difference  contributes to the discrepancy
of the two universes through a particular combination
$\g_n:=\<\overline{\g}_{ab}{\>_{}}_{n} + \frac{1}{2}\<\g {\>_{}}_{n}$
 (Eq.(\ref{eq:d_N-close3})).

The spectral scheme served as  a powerful tool for dealing with  the
present problem. It is effective to analyze the type of problems
involving  two or more geometries.
Indeed we were able to derive   several non-trivial relations describing the
discrepancy between the reality and its model which would
have been difficult to find out  without the spectral scheme.
Here we realize  primary importance of preparing
a suitable language for talking about an entangled situation before
tackling it.

We have observed that, as far as the linear approximation of geometry is valid,
$d_N ({\cal G},{\cal G}')$ does not increase with time prominently, rather
it shows the tendency to decrease.
This result is, at least, compatible with our belief in the validity of
the approximate description of the Universe in terms of a model.
 The present investigation is
 on just one pair of geometries within the linear-regime; it is now required to
 investigate  wider class of spacetimes as well as to study
 the behavior beyond the liner-regime along this line.

We have learned in this paper that it is practically possible
to proceed in this direction;
now we can improve  the
order of accuracy of approximations, e.g. the passage from
Eq.(\ref{eq:result}) to Eqs.(\ref{eq:final-result1})-(\ref{eq:final-result4}).
One drawback of the analysis linked
 with the linear structure-formation theory is that detailed
 features of the problem become, as it were,  hidden behind it. Therefore,
it is desirable  to analyze the same problem via another, more basic route,
i.e. by directly using  the fundamental equations
for $d_N$, $\dot{d}_N$, $\ddot{d}_N$ etc.~\cite{MS-evolution}.

As the next step of investigation, the
evolution of $d_N ({\cal G},{\cal G}')$ should be studied
beyond  the linear regime of the geometrical discrepancy.
It can be done since we have fundamental relations
Eqs.(\ref{eq:d_N}) along with
the spectral evolution equations~\cite{MS-evolution}.

Finally we note that our approach to the averaging problem in cosmology
is quite different from the ``usual" one which the term ``averaging" might
suggest. In the spectral scheme,
there is no ambiguous procedure of ``averaging geometry" such as $\<h_{ab}\>$
(recall the difficulty {\it (D1)} in \S {\ref{section:I}}).
Rather, we first define the concept of ``closeness" between geometries and
choose a model $\cal G$ which is a simple geometry and
reasonably close to the reality ${\cal G}'$
in the space of all spaces ${\cal S}_N$.
This mapping from ${\cal G}'$ to ${\cal G}$ is what we regard ``averaging"
in effect. Needless to say,  this procedure is mathematically well-defined.
Since the concept of ``closeness" ($d_N$) is scale-dependent and
apparatus-dependent, this mapping procedure (``averaging") is also
depends on both the observational scale we are interested in and
the apparatus we rely on. These are desirable properties and we can
expect that the framework outlined here is glimpsing
 a new way of viewing spaces as scale-dependent and apparatus-dependent
 objects.

The type of analysis developed here has a  general applicability to a wider
range of theories, not restricted to  cosmology.
It is an issue of the internal relation of  the triad
$(Reality,  Model,  Dynamics)$, which arises quite universally.
When  a theory deals with
objects characterized by
a sequence  of real numbers ordered in an increasing order
(``spectra" of the theory), $\{ \Lambda_n \}_{n=0,1,2,\cdots}$,
one can construct the
spectral measure of closeness similar to Eq.(\ref{eq:d_N_general}) along with
Eq.(\ref{eq:F_cond}). Then one can construct the spectral space $\cal S$,
which is similar to our ${\cal S}_N$. This space forms a metrizable space
when  the spectral measure of closeness (Eq.(\ref{eq:d_N_general}))
 is chosen suitably, e.g. like Eq.(\ref{eq:d_N}).\footnote{
 Stating more precisely, in order for the space $\cal S$ to possess
 desirable properties,  varieties of  objects in the theory should be
 rich enough; viz. for a given object $O$ there should
 always be another object $O'$ which is an infinitesimal modification of $O$,
 so that the spectra for
 $O$ and those for $O'$ are infinitesimally close to each
 other. This condition is satisfied in the case of ${\cal S}_N$; for any given
   geometry, there  always exists its infinitesimally modified
   geometry~\cite{MS-space}.}
 A ``good" theory should have such a property that
  two generic points close to each other in $\cal S$
 remain close during some period of time of concern.
 Then we are entitled to approximate  the reality by  a model or
 classify the objects into types.
The analysis presented here is the very first step for understanding the
{\it issue of types} in physics.
It is interesting to analyze  fundamental theories from this viewpoint.

\vskip 1cm

The author thanks the Ministry of Education and Science,
the Government of Japan and Inamori Foundation, Japan for financial support.

\makeatletter
\@addtoreset{equation}{section}
\def\theequation{\thesection\arabic{equation}}
\makeatother

\appendix

\begin{center}
\bf APPENDIX
\end{center}

\section{The basic modes}
\label{section:Appendix A}

Here we present the basic modes on the $t=constant$ spatial section,
$\Sigma_t$,
of the ``model" universe
$T^3 \times \mbox{\bf R}$ with Eq.(\ref{eq:metric-model}).
We recall that we adopt
the Cartesian-type spatial coordinates $x^a$ ($a=1,2,3$) which
take the value in $[0,1]$ with the identification $0 \sim 1$.
We note $\Delta = \frac{1}{a^2}\partial_a \partial_b \d^{ab}$.

Now, let
\[
       \cos_\s \theta :=
   \left\{
     \begin{array}{rl}
            \cos \theta & \  (\s=0)  \ \ ,   \\
            \sin \theta & \  (\s=1)  \ \ . \\
     \end{array}
   \right.
\]
 Then the eigenfunctions of $\Delta$ on $\Sigma_t$ can be represented as
\begin{eqnarray}
f^{(\vec{\s})}_{\vec{n}}(x_1,x_2,x_3)&=&
\frac{\sqrt{2^{(3-\#(\vec{\s}))}}}{\sqrt{a^3}}
\cos_{\s_1} 2\pi n_1 x_1\  \cos_{\s_2} 2\pi n_2 x_2\
 \cos_{\s_3} 2\pi n_3 x_3  \nonumber \\
& & (n_1, n_2, n_3 = 0,1,2, \cdots;\  \s_1,\s_2,\s_3 =0,1) \ \ ,
\label{eq:f_A}
\end{eqnarray}
where $\vec{\s}:=(\s_1,\s_2,\s_3)$,
$\vec{n}:=(n_1,n_2,n_3)$ and
 $\#(\vec{\s})$ denotes the number of zero elements
 in $\vec{\s}$.\footnote{See  the beginning of \S \ref{section:V}
 for more details on the notations such as $\#(\vec{\s})$.}
The corresponding spectra can be represented as
\begin{equation}
\l_{\vec{n}}=\frac{(2\pi)^2}{a^2}\vec{n} \cdot \vec{n}
            = \frac{(2\pi)^2}{a^2} (n_1^2 + n_2^2 + n_3^2)  \ \ .
\label{eq:lambda_A}
\end{equation}

We recall that we have introduced
the eigenfunction $\tf^{(\vec{\s})}_{\vec{n}}$  and
the eigenvalue $\tl_{\vec{n}}$ of
the Laplacian $\tilde{\Delta}$ on the regular 3-torus $T^3(1)$
(see after Eq.(\ref{eq:metric-model})). They are related to
 $f^{(\vec{\s})}_{\vec{n}}$ and  $\l_{\vec{n}}$ as
\begin{equation}
\tf^{(\vec{\s})}_{\vec{n}}:=\sqrt{a^3}\ f^{(\vec{\s})}_{\vec{n}}\ \ , \ \
\tl_{\vec{n}}:= a^2 \l_{\vec{n}} \ \ .
\label{eq:tilde_f}
\end{equation}

For notational simplicity,
we often  write $f_A$ and $\tf_A$ instead of $f^{(\vec{\s})}_{\vec{n}}$ and
$\tf^{(\vec{\s})}_{\vec{n}}$, respectively,   wherever
explicit indication of the label $(\vec{n}, \vec{\s})$ can be omitted.

We recall the conformal equivalence between $\Sigma_t$ and $T^3(1)$
(see the argument after Eq.(\ref{eq:metric-model})).
Because of  this equivalence,
we can define the integral over $T^3(1)$
for  any function  ${\cal A}$ and any
symmetric tensor field  ${\cal A}_{ab}$ on
$\Sigma_t$,
\begin{equation}
\< {\cal A} \tilde{{\>_{}}_A}:= \int_{T^3(1)}\ \tf_A \ {\cal A}
                                       \ \tf_A \  d^3x \ \ , \ \
\< {\cal A}_{ab} \tilde{{\>_{}}_A}
     := \frac{1}{\tl_A}\int_{T^3(1)} \
      \ \partial^a \tf_A \  {\cal A}_{ab} \
               \partial^b \tf_A \  d^3x  \ \ .
\label{eq:tilde_<A>}
\end{equation}
 Then, it is useful
 to introduce the coefficients $c^{(1)}_{AA'}$, $c^{(2)}_{AA'}$
and $c^{(3)}_{AA'}$ as
\begin{eqnarray}
c^{(1)}_{AA'}:&=& \< \tf_{A'} \tilde{{\>_{}}_A}
               = \sqrt{a^3} \< f_{A'} {\>_{}}_A   \nonumber \\
              &=&(-)^{\# (\vec{\s}_\|) }\frac{\sqrt{2}}{4}
               (\sqrt{2})^{\#(\vec{n}')} \d_{\vec{\s}',\vec{0}} \,
              \d_{\vec{n}',2 \vec{n}_\|}\, \d_{{\vec{n}'}_\perp, \vec{0}} \ \ ,
                                                         \nonumber \\
c^{(2)}_{AA'}:
     &=&\left\{
       \begin{array}{rl}
        \<\frac{1}{\tl_{A'}} \partial_a \partial_b \,
                \tf_{A'} \tilde{{\>_{}}_A} &
                = \sqrt{a^3} \<\frac{1}{\l_{A'}} \partial_a \partial_b \,
                     f_{A'} {\>_{}}_A \ \   (\vec{n}' \neq \vec{0}) \ \ ,  \\
           0 \ & \ \qquad \qquad (\vec{n}'=\vec{0}) \\
      \end{array}
     \right.
            \nonumber \\
   &&=   {\hat{n}_\|}^2 c^{(1)}_{AA'}
                                     \nonumber \\
c^{(3)}_{AA'}:&=& -\<\tf_{A'}\th_{ab}\tilde{{\>_{}}_A}
              = -\sqrt{a^3} \<f_{A'} h_{ab} {\>_{}}_A
                \nonumber \\
       &=&  (1-2{\hat{n}_\perp}^2)c^{(1)}_{AA'}
             \ \ .
\label{eq:c}
\end{eqnarray}
Here $\vec{n}_{\|}$  ($\vec{n}_{\perp}$) is, roughly speaking,
the component of $\vec{n}$   parallel (perpendicular) to $\vec{n}'$~\footnote{
See  the beginning of \S \ref{section:V} for notations.
}
; $\hat{n}_\|^2:= \left(\vec{n}_{\|}/|\vec{n}|\right)^2$ and
$\hat{n}_\perp^2:= \left(\vec{n}_{\perp}/|\vec{n}|\right)^2$, so that
 $\hat{n}_\|^2$ and  $\hat{n}_\perp^2$ possess only the information
 on the relative direction between $\vec{n}$ and $\vec{n}'$.
 Thus it is obvious that the coefficients
   $c^{(1)}_{AA'}$, $c^{(2)}_{AA'}$ and $c^{(3)}_{AA'}$ do not
depend on the information of $|\vec{n}|$, $|\vec{n}'|$ but
 on the information of the relative direction between
 $\vec{n}$ and $\vec{n}'$.

In particular, we note $c^{(1)}_{A0}=-c^{(3)}_{A0}=1$ and
 $c^{(2)}_{A0}=0$ by definition.
It is  also helpful to  note that, when $\vec{n}' \neq \vec{0}$,
\[
c^{(3)}_{AA'}
  =\<\frac{1}{\tl_{A'}}\tilde{\Delta} \tf_{A'} \th_{ab} \tilde{{\>_{}}_A}
          = \sqrt{a^3}\<\frac{1}{\l_{A'}}\tilde{\Delta}
               f_{A'} h_{ab} {\>_{}}_A \ \ .
\]
There is an identity
\begin{equation}
c^{(1)}_{AA'}-2c^{(2)}_{AA'}+c^{(3)}_{AA'}=0 \ \ ,
\label{eq:c-identity}
\end{equation}
so that one among  the three kinds of coefficients,
say $c^{(3)}_{AA'}$, can always be eliminated from the  formulas.

\subsection{Scalar modes}
\label{subsection:A-1}

From the basic modes $f_A$ in Eq.(\ref{eq:f_A}),
the components of the
perturbations   $\tg_{ab}$ in Eq.(\ref{eq:spatial-metric})
can be  produced.

Let us define $\ts_{ab}$ as
\[
\ts^A_{ab}:=\frac{1}{\tl_A}\partial_a \partial_b \tf_A
           - \frac{1}{3}{\tilde{\Delta}} \tf_A \th_{ab}\ \ .
\]
Then, $\tf_A \th_{ab}$ and $\ts^A_{ab}$ form the basis of
the scalar modes of the metric perturbation $\tg_{ab}$:
\begin{equation}
\tg_{ab}^{({\rm scalar})}= \frac{1}{3}\sum_A \gc_{\ A}\ \tf_A \ \th_{ab}
                    + {\sum_A}' \gs_{\ A}\  \ts^A_{ab}\ \ ,
\label{eq:scalar-perturbation}
\end{equation}
where $\gc_{\ A}$ and $\gs_{\ A}$ are arbitrary expansion coefficients.

We note some relations including  $\ts^A_{ab}$,
\begin{eqnarray}
\tilde{D}^a \ts^{A'}_{ab}&=&-\frac{2}{3}\partial_b \tf_{A'} \ \ , \ \
\tilde{D}^a \tilde{D}^b \ts^{A'}_{ab}
              = \frac{2}{3}\tl_{A'} \tf_{A'} \ \ , \nonumber \\
\tilde{\Delta} \ts^{A'}_{ab}
              &=&- \tl_{A'} \tf_{A'} \ \ ,     \nonumber  \\
\< \ts^{A'}_{ab} \tilde{{\>_{}}_A}
&=&c^{(2)}_{AA'}-\frac{1}{3}c^{(3)}_{AA'}
=\frac{1}{3}(c^{(1)}_{AA'}+c^{(2)}_{AA'})\ \ ,
\label{eq:sigma}
\end{eqnarray}
where Eq.(\ref{eq:c-identity}) has been used in the last line of
the fourth formula.

Then, it is straightforward to
compute from $h^{({\rm scalar})}_{ab}:=a^2(\d_{ab}+\tg_{ab}^{({\rm scalar})})$
 the curvature quantities up to $O(\g)$,
\begin{eqnarray}
\mbox{\boldmath $R$}^{({\rm scalar})}_{ab}&=&
      \frac{1}{6}{\sum_A}' (a^2 \l_A) (\gc_{\ A} + \gs_{\ A}) \ \tf_A \ \th_{ab}
      -\frac{1}{6}{\sum_A}'(\gc_{\ A} + \gs_{\ A})\partial_a \partial_b \tf_A
       \ \ , \nonumber \\
\mbox{\boldmath $R$}^{({\rm scalar})}&=&
      \frac{2}{3}{\sum_A}'  \l_A (\gc_{\ A} + \gs_{\ A}) \ \tf_A
       \ \ ,  \nonumber \\
r^{({\rm scalar})}_{ab} &=&
      -\frac{1}{6}{\sum_A}'(a^2 \l_A)(\gc_{\ A} + \gs_{\ A})\ts^A_{ab}\ \ .
\label{eq:curvature-scalar}
\end{eqnarray}

\subsection{Vector modes}
\label{subsection:A-2}

We choose the basic solutions of the eigenvalue problem,
 $\Delta v_{a}=-\l v_{a}$, with $D^a v_{a}=0$.
 We get the solution
\begin{equation}
v_a^{(\vec{n}, \vec{\s}, j)}
=a\ f_{(\vec{n}, \vec{\s})} \d_{\vec{n}\cdot \vec{e}^{(j)},0}
                      (e^{(j)})_a \ \ ,
\label{eq:vector-mode}
\end{equation}
where $(e^{(j)})_a:=\d_{ja}$ can be identified with
the unit vector in  $\mbox{\bf R}^3$ along $x^j$-axis ($j=1,2,3$), and
$\{ a (e^{(j)})^a\}_{j=1,2,3} $ forms the orthonormal bases
in the $t=constant$ geometry.\footnote{
Note that the indices are
lowered by $h_{ab}$ in Eq.(\ref{eq:metric-model}) and raised by
its inverse, $h^{ab}$.
}
 They are  normalized as
\[
\int v_a^{(\vec{n}, \vec{\s}, j)} h^{ab} v_b^{(\vec{n}', \vec{\s}', j')}
= \d_{j,j'}\d_{\vec{n},\vec{n}'} \d_{\vec{\s},\vec{\s}'} \ \ .
\]

From $v_a^{(\vec{n}, \vec{\s}, j)}$, we can construct
the traceless tensor $\zeta_{ab}^{(\vec{n}, \vec{\s}, j)}
=\frac{1}{\sqrt{2\l_{\vec{n}}}}(D_a v_b + D_b v_a)$, which is
normalized as
$(\zeta_{ab}^{(\vec{n}, \vec{\s}, j)}, \zeta_{cd}^{(\vec{n}', \vec{\s}', j')})
= \d_{j,j'} \d_{\vec{n},\vec{n}'} \d_{\vec{\s},\vec{\s}'}$.

Then, $\tilde{\zeta}_{ab}^{(A,j)}
:=\sqrt{a}\zeta_{ab}^{(A,j)}$ form the basis of
the vector modes of the metric perturbation $\tg_{ab}$:
\begin{equation}
\tg_{ab}^{({\rm vector})}
= {\sum_{A,j}}'  \gv_{\ Aj}\ \tilde{\zeta}^{Aj}_{ab}\ \ .
\label{eq:vector-perturbation}
\end{equation}
Wherever the detailed index structure is not essential, let us
write $A$ instead of $Aj$ for notational neatness,
like $\gv_A$, $\tilde{\zeta}^A_{ab}$ and
${\sum'}_A  \gv_{\ A}\ \tilde{\zeta}^{A}_{ab}$.
One can show that
\begin{equation}
\< \tilde{\zeta}^{A'j}_{ab} \tilde{{\>_{}}_A}=0\ \ .
\label{eq:<zeta>}
\end{equation}

The curvature quantities up to $O(\g)$ turn out to be
\begin{equation}
\mbox{\boldmath $R$}^{({\rm vector})}_{ab} =  0 \ \ , \ \
\mbox{\boldmath $R$}^{({\rm vector})} = 0 \ \ , \ \
r^{({\rm vector})}_{ab} = 0\ \ .
\label{eq:curvature-vector}
\end{equation}

\subsection{Tensor modes}
\label{subsection:A-3}

We choose the basic solutions of the eigenvalue problem,
 $\Delta w_{ab}=-\l w_{ab}$, with $D^a w_{ab}=0$,
 $w_a^{\ a}=0$.
 We get the solution
\begin{equation}
w_{ab}^{(\vec{n}, \vec{\s}, \a)}
=a^2 \d_{\vec{n}_{\perp (\vec{e}^{(\a)})} , \vec{0} }\
   \d_{\vec{\s}_\perp (\vec{e}^{(\a)}) , \vec{0}}\
\ f_{(\vec{n}, \vec{\s})}\  \nu_{ab}^{(\a)}\ \ ,
\label{eq:tensor-mode}
\end{equation}
where
\begin{eqnarray}
\nu_{ab}^{(1)}&=&
   \frac{1}{\sqrt{2}}
   \left(
   \begin{array}{ccc}
   0 &  &   \\
    & 1 &    \\
    &  &  -1
   \end{array}
   \right), \ \
\nu_{ab}^{(2)}=
   \frac{1}{\sqrt{2}}
   \left(
   \begin{array}{ccc}
   -1 &  &   \\
    & 0  &    \\
    &    &  1
   \end{array}
   \right), \ \
\nu_{ab}^{(3)}=
   \frac{1}{\sqrt{2}}
   \left(
   \begin{array}{ccc}
    1 &     &    \\
      & -1  &    \\
      &     &  0
   \end{array}
   \right),   \nonumber \\
\nu_{ab}^{(I)}&=&
   \frac{1}{\sqrt{2}}
   \left(
   \begin{array}{ccc}
    &   &   \\
    &   & 1   \\
    & 1 &
   \end{array}
   \right), \ \
\nu_{ab}^{(II)}=
   \frac{1}{\sqrt{2}}
   \left(
   \begin{array}{ccc}
     &  & 1  \\
     &  &    \\
   1 &  &
   \end{array}
   \right), \ \
\nu_{ab}^{(III)}=
   \frac{1}{\sqrt{2}}
   \left(
   \begin{array}{ccc}
       & 1 &    \\
     1 &   &    \\
       &   &
   \end{array}
   \right)\ ,
\label{eq:nu-matrix}
\end{eqnarray}
and $\vec{e}^{(1)}=\vec{e}^{(I)}=(1,0,0)$,
$\vec{e}^{(2)}=\vec{e}^{(II)}=(0,1,0)$,
$\vec{e}^{(3)}=\vec{e}^{(III)}=(0,0,1)$.

 They are  normalized as
\[
\int w_{ac}^{(\vec{n}, \vec{\s}, \a)}
      h^{ab}h^{cd} w_{bd}^{(\vec{n}', \vec{\s}', \a')}
= \d_{\a,\a'}\d_{\vec{n},\vec{n}'} \d_{\vec{\s},\vec{\s}'}\ \ .
\]

Then, $\tw_{ab}^{(A,\a)}
:=\frac{1}{\sqrt{a}}w_{ab}^{(A,\a)}$ form the basis of
the vector modes of the metric perturbation $\tg_{ab}$:
\begin{equation}
\tg_{ab}^{({\rm tensor})}
= {\sum_{A,\a}}  \gt_{\ A\a}\ \tw^{A\a}_{ab}\ \ .
\label{eq:tensor-perturbation}
\end{equation}
Let us introduce the coefficient $c^{(T)}_{AA'\a}$ as
\begin{equation}
c^{(T)}_{AA'\a} := \< \tw^{A' \a}_{ab} \tilde{{\>_{}}}_{A}\ \ .
\label{eq:c^T1}
\end{equation}

For $\a=i=1,2,3$,
\begin{equation}
c^{(T)}_{(\vec{n},\vec{\s})(\vec{m},\vec{\s}')i}
=\left\{
       \begin{array}{c}
(1-\d_{n_i,0})\frac{(2\pi)^2}{\tilde{\l}_{\vec{n}}}
(n_{i+1}^2 - n_{i+2}^2)c^{(1)}_{(\vec{n},\vec{\s})(\vec{m},\vec{\s}')} \\
     ({\rm when \ at\  most\   one\  of\ } n_1,n_2,n_3\
                  {\rm vanishes} ) \ \ , \\
  0 \qquad \qquad \qquad \qquad \qquad \qquad  \\
   ({\rm when\ at\  least\  two\   of\ } n_1,n_2,n_3\ {\rm vanish} )\ \ . \\
      \end{array}
     \right.
\label{eq:c^T2}
\end{equation}
For $\a=I,II,III$, it turns out to be $c^{(T)}_{AA'\a}=0$.

The curvature quantities up to $O(\g)$ become
\begin{equation}
\mbox{\boldmath $R$}^{({\rm tensor})}_{ab} =
\frac{1}{2}{\sum_A}'\l_A\gt_A \tw^A_{ab} \ \ ,  \ \
\mbox{\boldmath $R$}^{({\rm tensor})} = 0 \ \ ,  \ \
r^{({\rm tensor})}_{ab} =
\frac{1}{2}{\sum_A}'\l_A\gt_A \tw^A_{ab} \ \ ,
\label{eq:curvature-tensor}
\end{equation}
where we have used a simplified index notation $A$ for $A\a$
for notational neatness.
We follow this convention wherever the detailed index structure
can be omitted, and we write simply, e.g., $\gt_A$, $\tw^A_{ab}$ and
${\sum_{A}}  \gt_A\ \tw^{A}_{ab}$.

\section{Basic results of the linear structure-formation theory}
\label{section:Appendix B}

We summarize the time evolution of the linear perturbations
 for the typical cases
  when (I) $\l_{\rm phys} \gg c_s t$ and $\theta \neq 0$,
(I') $\l_{\rm phys} \gg c_s t$ and $\theta = 0$ and
(II) $\l_{\rm phys} \ll c_s t$, where $\l_{\rm phys}$ is the physical
wave length of the perturbation mode~\cite{PJEP}.

\subsection{Behavior of perturbations:
The case of $\l_{\rm phys} \gg c_s t$ and $\theta \neq 0$}
\label{subsection:B-1}

A perturbation mode satisfying   $\l_{\rm phys} \gg c_s t$
 does not oscillate sufficiently  during the cosmological time-scale $t$.
  It is a long-wave length, slowly varying fluctuation compared to $t$.
Let $\tau$ be the oscillation time-scale of the perturbation,
$\tau = \l_{\rm phys}/c_s$.
Then $t \ll \tau$, so that
the only important time-scale in this situation is $t$,
and perturbations naturally  vary as a power of $t$.

We assume $\nu =constant = c_s^2$ for simplicity.

In this case,
\begin{eqnarray}
\d(\vec{x},t)&=& F(\vec{x})\ t^{\frac{9\nu-1}{3(1+\nu)}}\ \ , \ \
 \g(\vec{x},t) = -\frac{2(1+3\nu)}{\nu (9\nu -1)} \d \ \ , \nonumber \\
 \theta(\vec{x},t) &=& \frac{(1-\nu)(6\nu +1)}{3\nu(1+\nu)^2} \frac{\d}{t}\ \
                   \propto \frac{1}{\nu}t^{-\frac{2(2-3\nu)}{3(1+\nu)}}\ \ .
\label{eq:result1}
\end{eqnarray}
Here $F(\vec{x})$ is an arbitrary function determined by the
initial condition imposed at a certain time $t=t_0$;
the factor $1/\nu$ has been shown in the last term of the last line
to indicate the singular behavior for $\nu \downarrow 0$.

As a typical situation, it is illustrative to  consider the plane-wave type
component of perturbation which is moving in the $x^3$-direction.
In this case, the behavior of $\tg_{33}$ and $\tg_{11}\sim \tg_{22}$
are
\begin{equation}
\tg_{33} \sim -2 \tg_{11} \sim -\frac{8(6\nu+1)}{9\nu(1+\nu)^2}
                               \frac{a^2}{k^2 t^2}\ \d
                      \propto \frac{1}{\nu}
                      \left( \frac{\l_{\rm phys}}{c_s t}\right)^2 \d
                      \propto \frac{1}{\nu k^2}t^{-\frac{1-\nu}{1+\nu}} \ \ ,
\label{eq:result-gamma1}
\end{equation}
where $k$ is the coordinate wave-number ($k=2 \pi a /\l_{\rm phys}$).

It is convenient to summarize the relation of the order of magnitude of
several quantities in a symbolical manner as
\begin{equation}
(|\g| \sim |\d| \sim |t\theta|) \sim (|t \dot{\g}| \sim |t \dot{\d}| \sim
|t^2 \dot{\theta}|)
\ll |\tg_{kl}| \sim |t \dot{\tg}_{kl}| \ \ .
\label{eq:result-order1}
\end{equation}
Regarding the rate of change of $\tg_{11}$, we see that
$t|\dot{\tg}_{11}|\sim |\tg_{11}|$. It means that
$|\tg_{kl}|$ decreases prominently within the cosmological time-scale.
It implies that $|\d (t_{in})| \sim \left( \frac{c_s t}{\l_{\rm phys}}\right)^2
|\tg_{11}(t_{in})|$$\sim |\tg_{11}(t_{in})|$$ \ll |\tg_{11}(t_0)|$ where
$t_{in} $ is the time when the perturbation of scale $\l_{\rm phys}$
``enters within the horizon" due to the enlargement of
the horizon scale ($\sim ct$), viz. $\l_{\rm phys} \sim c_s t$.
It indicates the linear perturbation theory is valid even after
$t= t_{in}$~\cite{PJEP}.

\subsection{Behavior of perturbations:
The case of $\l_{\rm phys} \gg c_s t$ and $\theta = 0$}
\label{subsection:B-2}

In this case,
\begin{equation}
\d(\vec{x},t) = F(\vec{x})\ t^{\frac{2(1+3\nu)}{3(1+\nu)}}\ \ , \ \
 \g(\vec{x},t) = -\frac{2}{1+\nu} \d \ \ .
\label{eq:result2}
\end{equation}

For the plane-wave type
component of perturbation  moving in the $x^3$-direction,
the behavior of $\tg_{33}$ and $\tg_{11}\sim \tg_{22}$
becomes
\begin{equation}
\tg_{33} \sim -2 \tg_{11} \sim -\frac{8(9\nu+5)}{9(1+\nu)^3}
                               \frac{a^2}{k^2 t^2}\ \d
                               \propto
                               \left( \frac{\l_{\rm phys}}{c_s t}\right)^2 \d
                               \propto \frac{1}{k^2} t^0 \ \ .
\label{eq:result-gamma2}
\end{equation}
On the other hand, it turns out to be
\begin{equation}
t\dot{\tg}_{33} \sim -2 t\dot{\tg}_{11}
     = \frac{8\nu}{(1+\nu)(5+3\nu)} \d\ \ .
\label{eq:result-gamma-dot2}
\end{equation}

We can summarize the relation of the order of magnitude of
several quantities symbolically as
\begin{equation}
  (|\g| \sim |\d|) \sim (|t \dot{\g}| \sim  |t \dot{\tg}_{kl}|
  \sim |t \dot{\d}|)
  \ll |\tg_{kl}| \ \ ,
\label{eq:result-order2}
\end{equation}

The peculiar feature of the case  $\theta = 0$ compared to the case
$\theta \neq 0$ is that $|\tg_{kl}|$ does not change
prominently within the cosmological time-scale
($t|\dot{\tg}_{11}|\sim |\d| \ll |\tg_{11}| $).

\subsection{Behavior of perturbations:
The case of $\l_{\rm phys} \ll c_s t$}
\label{subsection:B-3}

A perturbation mode satisfying   $\l_{\rm phys} \ll c_s t$
  oscillates sufficiently  during the cosmological time-scale $t$.
  It is a short-wave length, fast varying fluctuation compared to $t$.
Now $t \gg \tau$
($\tau=\frac{\l_{\rm phys}}{c_s}=\frac{2\pi a}{k c_s}$
characterizes the  oscillation period), so that  we naturally
pay attention to the process whose time-scale is $O(\tau)$.

In this case, various quantities oscillate with the time-scale $\tau$:
\begin{eqnarray}
\d & \simeq & - \g \simeq -\tg_{33} \propto
   \left( \frac{\rho+p}{k c_s a^4 \rho^2} \right)^{1/2}
   \exp -i\int^t \frac{kc_s(t')}{a(t')}dt'\ \ , \nonumber \\
    \theta & \simeq & -\frac{1}{1+\nu} \dot{\d}  \ \ , \nonumber \\
|\d| & \simeq & |\g| \sim |\tg_{33}| \sim \tau |\theta| \propto
   \left( \frac{\rho+p}{k c_s a^4 \rho^2} \right)^{1/2}
    \propto \frac{1}{\sqrt{k}}  a^{-\frac{1-3\nu}{2}}
   \propto \frac{1}{\sqrt{k}} t^{-\frac{1-3\nu}{3(1+\nu)}} \ \ ,
\label{eq:result3}
\end{eqnarray}
where we have assumed $\rho \sim \rho_{\rm av} \propto
a^{-3(1+\nu)}\propto t^{-2}$.

We can summarize  the relation of the order of magnitude of
several quantities in a symbolical manner as
(``$\sim$" indicates that the difference between the both-hand sides is
at most $o(\d)$.)
\begin{eqnarray}
(|\d| & \sim & |\g| \sim |\tg_{33}|)
  \sim (\tau |\dot{\d}| \sim \tau |\theta| )
\ll \frac{\l_{\rm phys}}{c_s t} \ll 1 \ \ , \nonumber \\
|\tg_{11}| &\sim & |\tg_{22}| \sim 0\  (o(\d)) \ \ , \nonumber \\
(\tau |\dot{\g}| & \sim & \tau |\dot{\tg}_{kl}|) \sim
       \left( \frac{\l_{\rm phys}}{c_s t} \right)^2 |\d| \ll |\d| \ \ .
\label{eq:result-order3}
\end{eqnarray}



\begin{thebibliography}{99}
\bibitem{AVE}
For instance, G.F.R. Ellis, in {\sl ``Proceedings of the Tenth
International Conference on General Relativity and Gravitation''},
edited by B. Bertotti, F. De Felice and A. Pascolini
(Reidel, Dordrecht, 1984); H. Sato, {\it ibid}.
\bibitem{AVE2}
 See also,
A. Krasi{\'n}ski, {\sl Inhomogeneous Cosmological Models}
(Cambridge University Press, Cambridge, 1997), Chapter 8,  and the
references therein.
\bibitem{MS-spectral}
M. Seriu, Phys. Rev. D{\bf 53}, 6902 (1996).
\bibitem{MS-space}
M. Seriu, Comm. Math. Phys. {\bf 209}, 393 (2000).
\bibitem{MS-evolution}
M. Seriu, Phys. Rev. {\bf D62} \#023516 (2000).
\bibitem{MS-AVE}
M. Seriu,  Gen. Rel. Grav. {\bf 32}, 1473 (2000).
\bibitem{MS-JGRG}
M. Seriu, in {\sl ``Proceedings of the 8th Workshop
 on General Relativity and Gravitation"}
  (K.Oohara et. al. (eds.), Niigata University, 1999), 334.
\bibitem{MS-scale}
M. Seriu, Phys. Let. B{\bf 319}, 74 (1993).
\bibitem{Numerical}
E.g., K. T. Inoue, Class. Quantum. Grav. {\bf 16}, 3071 (1999).
\bibitem{CH}
See e.g., I. Chavel,  {\sl Eigenvalues in Riemannian Geometry}
(Academic Press, Orland,  1984).
\bibitem{Kac}
M. Kac, Am. Math. Mon. {\bf 73}(4), 1 (1966).
\bibitem{KE}
J.L. Kelly,  {\sl General Topology}
 (D. van Nostrand, Princeton, 1955).
\bibitem{YA}
K. Yano,  {\sl  Metric Spaces and Topological Structures}
  (Kyoritsu Shuppan, Tokyo, 1998), Chapters 3 and  4.
\bibitem{PJEP}
For a standard treatment of the structure-formation theory, see
P.J.E. Peebles, {\sl The Large-Scale Structure of the Universe}
(Princeton University Press, Princeton, 1980), Chapter 5.
\end{thebibliography}
\end{document}